\documentstyle[pre,aps,epsf]{revtex}

\newcommand{\bq}{\begin{equation}}
\newcommand{\ee}{\end{equation}}
\newcommand{\fr}[2]{\frac{#1}{#2}}

\begin{document}
\draft




\title{Coherent propagation of
interacting particles in a random potential: the Mechanism of
enhancement }

\author{I. V. Ponomarev,$^{1,2}$}

\author{P. G. Silvestrov$^{1,3}$}

\address{$^1$ Budker Institute of Nuclear Physics, 630090 Novosibirsk,
Russia}

\address{$^2$ School of Physics, The University of New South Wales,
Sydney 2052, Australia}

\address{$^3$ The Niels Bohr Institute, Blegdamsvej 17, DK-2100
Copenhagen {\O}, Denmark}

\maketitle


\begin{abstract}
Coherent propagation of two interacting particles in $1d$ weak random
potential is considered. An accurate estimate of the matrix element of
interaction in the basis of localized states leads to mapping onto the
relevant matrix model. This mapping allows to clarify the mechanism of
enhancement of the localization length which turns out to be rather
different from the one considered in the literature.  Although the
existence of enhancement is transparent, an analytical solution of the
matrix model was found only for very short samples. For a more
realistic situation numerical simulations were performed. The result
of these simulations is consistent with
\[
l_{2}/l_1 \sim l_1^{\gamma} \ \ ,
\]
where $l_1$ and $l_2$ are the single and two particle localization
lengths and the exponent $\gamma$ depends on the strength of the
interaction. In particular, in the limit of strong particle--particle
interaction there is no enhancement of the coherent propagation at all
($l_{2} \approx l_1$).

\end{abstract}

\pacs{PACS numbers: 72.15.Rn, 71.30.+h, 05.45.+b}

\section{Introduction}\label{sec:1}

The enhancement of the propagation length for two interacting
particles (TIP) in a one-dimensional random potential was predicted in
a paper by D.~Shepelyansky \cite{Dima} a couple of years ago. This
result has attracted broad interest and stimulated active analytical
\cite{Imry,Fyodorov,Frahm,Pichard3,Sushkov} and numerical
\cite{Pichard,Pichard2,Oppen,Dima3,Pichard4}
investigations. 2-3-dimensional and quasi-$1d$ extensions of the model
\cite{Dima}, as well as many other related problems have also been
considered in various papers (see
e.g. \cite{Imry,Pichard3,Dima4}). Moreover some of these new results
may even be better established than the original one \cite{Dima} (see
also \cite{Dorokhov}).

More specifically, the first estimate \cite{Dima} of the two-particle
localization length was
\bq\label{l2}
l_2 \sim l_1^2 \ \ ,
\ee
where $l_1$ and $l_2$ are the one and two particle localization
lengths.

Numerical simulations first of all have confirmed the existence of
enhancement. As for the specific form of $l_2$, some authors
\cite{Pichard} see deviations from (\ref{l2}), whereas others
\cite{Oppen} report on the agreement with (\ref{l2}) (more concretely
the authors of \cite{Oppen} agree with \cite{Dima} in the $l_1$
dependence of $l_2$ although they still disagree in the overall
normalization constant in (\ref{l2})). However, the actual enhancement
$l_2/l_1$ which was observed in numerical simulations, turns out to be
rather small and varies from $l_2/l_1 \sim 2$ in the first paper
\cite{Dima} to $l_2/l_1 \sim 3-6$ for the most advanced computations
\cite{Oppen}.

We are going to consider particles moving in an exactly $1d$, weak,
random potential. This means that the Anderson localization length
$l_1$ is much larger than the de~Broglie wave length $\lambda$ (for
the Anderson hopping Hamiltonian (\ref{H1}) below $\lambda\sim 1$ and
$l_1\gg 1$). Let the total length of the sample be $L\gg l_1$. The
main achievement of \cite{Dima} was the prediction of the existence of
very unusual bound states for TIP (we will call them coherent states
in order to distinguish from the usual molecular states
\cite{molecules}). The typical distance between particles for these
new states is rather large, $x_1-x_2 \sim l_1$, but the joint
propagation length for two particles turns out to be even much larger 
$\Delta (x_1+x_2) \sim l_2 \gg l_1$. The total number of these new
eigenstates is also sufficiently large: $\Delta N_c\sim
(L/\lambda)(l_1/\lambda)$.

The central point for all the methods applied to calculation of the
coherent propagation length $l_2$ was an estimate of the matrix
element of the interaction in the basis of products of single particle
localized states. However, as we will show in the following section,
the estimate of the matrix element in ref.  \cite{Dima} crucially
depends on the oversimplified assumption regarding the behavior of the
single particle wave function and therefore is
irrelevant. Surprisingly, all the authors of the following papers have
accepted the matrix element estimate of \cite{Dima} without any
critical analysis.

This is why the main part of our paper will be devoted to an accurate
estimate of the matrix element. This estimate allows us to perform
mapping of the two-particle problem onto the physically relevant
matrix model.

Surely the mapping itself is impossible without some (properly
motivated) assumptions and order of magnitude estimates. Moreover,
although we see a mechanism which should lead to the enhancement of
the coherent propagation length, we have no rigorous proof that one
could not find another source of enhancement. Therefore in the light
of contradiction with existing predictions it should be very useful to
find at least one new rigorous result. To this end we consider in
Section 3 the strong coupling limit of the Shepelyansky model. It will
be shown without any assumptions that for the very strong interaction
between particles all the enhancement disappears and $l_2 = l_1$.
Thus the problem of two interacting particles exhibits some kind of
duality between the parameters $U/t$ for weak ($U\ll t$) and $t/U$ for
strong ($t\ll U$) coupling cases \cite{Pichthanks}, where $U$ is the
strength of interaction and $t$ is the typical single-particle energy
(\ref{H1},\ref{Vint}). In both weak and strong coupling limits the
enhancement disappears. As a function of $U$ the coherent propagation
length $l_2$ should reach the maximum at some value of $U/t\sim 1$.

As we will show in the following section, the two main features of the
matrix element of interaction lead us to associate the TIP problem
with the matrix models which differs strongly from the previously
considered ones \cite{Dima,Fyodorov,Frahm,Pichard3}. First of all it
is the hidden hierarchy of the matrix elements. In general the matrix
element of interaction in the basis of noninteracting two-particle
states turns out to be much smaller than it was expected from the
first estimate of ref. \cite{Dima}. Only very small part of the matrix
elements (namely $\sim 1/\sqrt{l_1}$) may exceed the original estimate
by Shepelyansky \cite{Dima}. This will be may be the most surprising
result of our paper that such a few large matrix elements still may
lead to a considerable enhancement. Another important feature of the
matrix elements is the hidden order in their distribution. As we will
show in  Sections 5,6, after the proper ordering of the
noninteracting basis the complicated hierarchy of the matrix elements
may be described by the corresponding enveloping function of the
relevant banded Gaussian random matrix model.

The matrix model we are going to consider turns out to be much more
complicated than those investigated in
\cite{Dima,Pichard,Pichard2,Oppen,Dima3,Fyodorov,Frahm,Pichard4}.
Therefore at first it would be useful to consider a simplified version
of the problem. To this end in  Section 5 we consider the TIP in
the short sample with the total size $L$ being of the same order of
magnitude as the single particle localization length $L\sim l_1$. In
this case the corresponding random matrix is the so called Power-law
Random Band Matrix (PRBM). The elements of this matrix decrease in a
power-law fashion $M_{ij}\sim |i-j|^{-1}$ as one goes farther from the
diagonal. Among the PRBM matrices of the general form with $M_{ij}\sim
|i-j|^{-\alpha}$ those with $\alpha=1$ correspond to the phase
transition from localized to delocalized regime \cite{mirlin}. For
short sample we consider the so called inverse participation ratio
$l_{ipr}$ (which is effectively the number of noninteracting
two-particle states mixed into one chaotic eigenstate). The result
reads
\bq\label{lpr}
 l_{ipr} \sim l_1^{\gamma_{ipr}} \ \ , 
\ee
where $\gamma_{ipr}$ is a function of the strength of the interaction
approaching the value $\gamma_{ipr}=0$ in the weak and strong coupling
limit. Equation (\ref{lpr}) is proved analytically by the
Renormalization Group method for $\gamma_{ipr} \ll 1$ but still for
$\gamma_{ipr}\ln(l_1) \gg 1$. The basic idea of calculating
(\ref{lpr}) follows the method of Levitov \cite{levitov} who
considered an even more complicated $3d$ problem. Numerical
simulations also support the result (\ref{lpr}) for $\gamma_{ipr} \sim
1$.

Unfortunately, for the realistic model with $L \gg l_1$ we could not
find any convincing analytical solution (the matrix model itself will
be described in Section 6). Nevertheless, we at least see that for a
large sample the chaotic mixing of noninteracting two-particle states
is systematically enhanced compared to a short sample. For example, it
may be the same expression (\ref{lpr}) for $ l_{ipr}$ but with
$\gamma' > \gamma$. The perturbation theory for large $L$ may be used
only if $ \langle l_{ipr}\rangle -1 \ll 1$. Nevertheless the physical
expansion parameter for this perturbation theory still is $
\ln(l_1)U/t$ for the weak coupling and $ \ln(l_1)t/U$ for the strong
coupling limit.

Therefore we have to perform numerical simulations for the matrix
model associated with two particles on a large sample. The details of
numerical procedure will be considered in Section 6 and now we will
only make some general comments. The inverse participation ratio
$l_{ipr}$ characterizes the complexity of the typical eigenstate in
the Hilbert space formed by noninteracting two-particle states. On the
other hand the coherent propagation length $l_2$ directly measures the
spatial (lattice sites) dimension of the corresponding wave
function. For practical calculations (Section 6) we define $l_2$ throw
the mean squared size of the wave packet in longitudinal
direction. We present our final result in the form
\bq\label{l2new}
l_{2}/l_1 \sim l_1^{\gamma} \ \ .
\ee

First of all, {\it we definitely see that the exponent $\gamma$ is a
function of the strength of interaction}. Unfortunately we do not know
too much about this dependence of $\gamma$ on $U$. The only claim we
can make is that $\gamma$ should go to zero for the weak ($U\ll t$)
and strong ($t\ll U$) coupling limiting cases (see also
\cite{discussion}). For arbitrary $U\sim t$ we expect $\gamma \sim
1$. For the concrete choice of parameters which we have considered
numerically it was also $\gamma < 1$. However, we do not know whether
the same inequality $\gamma < 1$ holds for any strength of
interaction. Also for $l_1 \gg \lambda$ the exponent $\gamma$ in 
(\ref{l2new}) does not depend strongly on the strength of disorder
or, in other words, $\gamma$ does not depend on $l_1$. The 
two-particle localization length of the form 
(\ref{l2new}) with $\gamma< 1$ was found also numerically in 
\cite{Pichard} by the transfer matrix method.

Of course the expression (\ref{l2new}) (as well as (\ref{lpr})) should
be used only for large $l_1$. This means that if one tries to extend
(\ref{l2new}) to small $l_1$, $\gamma$ will become a function of $l_1$
as well. For example, in order to fit the results of simulations, we
use $\gamma = \gamma_0 +c/l_1$. Our data (see figs. 2,3 below) shows
that at least for the coherent propagation length $l_2$ the value of
$\gamma$ does not vary strongly with $l_1$ . However, our numerical
accuracy still is not enough to exclude completely some exotic
dependence of $\gamma$ on $l_1$ for large $l_1$, say $\gamma = a
+b\ln(l_1)$ .

This paper is organized as follows. In Section 2 we give a general
formulation of the problem and make a rough estimate of the matrix
element of the interaction. It is shown by considering the modified
Thouless block picture \cite{Thouless,Imry} that effectively the
interaction between particles is enhanced by a factor $\sim 
\ln(l_1)$. In principle the material of Section 2 allows one to
perform the mapping onto the matrix model which is done in Sections
5,6. However, in the next two sections we try to build a more stable
foundation for this mapping. In Section 3 we consider the strong
coupling limit of the Shepelyansky model. This consideration provides
us with a better understanding as to how to distinguish regular and
chaotic effects due to inter-particle interaction. On the other hand,
the exact solution of the two-particle problem in the strong coupling
regime allows us to perform the critical revision of the existing
estimates of the coherent propagation length. In Section 4 we present
a more rigorous estimate of the interaction matrix element than the
one in Section 2. By averaging over the disordered potential in the
``unimportant'' part of its Fourier spectrum, which is not responsible
for the global features of the localized single-particle states, 
we get rid of the
problem of rapid oscillations in the matrix element. In Section 5 we
investigate the effect of interaction in the short sample. Within the
Renormalization Group approach of \cite{levitov} a nontrivial solution
of the model is found analytically at least for the weak effective
interaction case. Finally, in the last Section 6 we describe the
mapping of our coherent propagation problem onto the eigenvalue
problem for some special Random Band Matrices. Features of this last
model are investigated numerically.

\section{Formulation of the problem and preliminary 
estimates}\label{sec:2}

Following \cite{Dima} consider  two particles on a $1d$
lattice with the Hamiltonian
\bq\label{Ham}
H_{tot} = H(n_1) +H(n_2) +V_{int} \ \ .
\ee
There are no serious contradictions against considering the same
problem in continuous $1d$-space. The Anderson lattice Hamiltonian,
which is used traditionally, only simplifies the numerical
calculations.  The single particle Hamiltonian has the following
nonzero elements for transitions between $n$-th and $m$-th sites of
the lattice
\bq\label{H1}
H_{nm}= -t(\delta_{n,m+1}+\delta_{n,m-1}) +w_n \delta_{n,m} \
\ .
\ee
Here $w_n$ is the random Gaussian potential and we suppose
that the disorder is weak compared to the kinetic energy:
\bq\label{weak}
\overline{ w_n w_m} \equiv w^2 \delta_{nm} \ \ \ , \ \ \ 
w \ll t \ \ .
\ee
In many papers devoted to the TIP
delocalization problem the disordered potential is usually chosen to
be uniformly distributed within the range $-W/2 <w_{n} <W/2$ . The
comparison with our results in this case may be made via trivial
replacement $w^2 =W^2/12$ .  The Hubbard on-site interaction is
defined as
\bq\label{Vint}
\bigl( V_{int} \psi \bigr)_{n_1,n_2} = U \delta_{n_1,n_2}
\psi(n_1,n_2) \ \ .
\ee
 
In general we assume that the interaction strength $U$ is of the same
order of magnitude as the hopping matrix element $t$, though the more
or less clear analytical results may be obtained only in the weak $(U
\ll t)$ and strong $(U \gg t)$ coupling limits.

For simplicity we consider  distinguishable particles, but all our
results are equally valid for bosons or fermions with opposite spins.

The Anderson localization length has the form \cite{MacKinnon}
\bq\label{l1}
l_1 = 2 \fr{4t^2-\epsilon^2}{w^2} \ \ ,
\ee
where $\epsilon$ is the single particle energy ($-2t < \epsilon <
2t$).  We will not be interested in the edges of the energy
zone. Therefore for our purposes it will be enough to remember that
$l_1 \approx (t/w)^2$. For the lattice model (\ref{Ham},\ref{H1}) both
$l_1$ and $l_2$ are naturally dimensionless.

An important feature of the single particle Hamiltonian (\ref{H1}),
which was completely ignored in \cite{Dima}, is that due to the
weakness of disorder it almost conserves the momentum. It is natural
to parameterize the single particle energy $\epsilon$ by the momentum
$k$ (again we are not interested in the very edges of the spectrum)
\bq\label{eps}
\epsilon = -2t \cos k \ \ .
\ee

Because of $w \ll t$, the eigenfunction of (\ref{H1})
$\psi_{\epsilon}(n)$ should be a linear combination of $\cos
(k_{\epsilon}n)$ and $\sin (k_{\epsilon}n)$ with slowly varying
amplitudes within the intervals small compared to $l_1$
(\ref{l1}). Therefore, it is convenient to consider the plane wave
basis
\bq\label{pwbasis}
\psi_{\epsilon}(n) = \sum_k C(k) |k\rangle \ \ ,  \ \ |k\rangle =
\fr{1}{\sqrt{L}} e^{ikn} \ \ ,
\ee
where $L$ is the total size of the sample $(L \gg l_1)$. Now the
amplitude $C(k)$ has  two narrow peaks $(\Delta k \sim 1/l_1)$
around $k=\pm k_{\epsilon}$.

Being the Fourier transform of the localized oscillating function
$\psi_{\epsilon}(n)$, the amplitude $C(k)$ should manifest some simple
features.  Let the function $\psi_{\epsilon}(n)$ be localized around
$n=n_0$ . Then one has
\begin{eqnarray}\label{n0}
 C({-k}) = C({k})^* \  &,& \nonumber \\
C({k}) = e^{ikn_0} \tilde{C}(k)  \ 
&,& \\
\tilde{C}(k) = f((k-k_{\epsilon})l_1)  \ &,& \nonumber
\end{eqnarray}
where $f(x)$ is some  smooth function concentrated around the
region $|x|\sim 1$.

The ensemble averaged value of  $|C(k)|^2$ may be also extracted
from the textbook \cite{Pastur}
\bq\label{gredesk}
\langle |C(k)|^2 \rangle \sim 
\left[
(\epsilon -\epsilon_{\scriptscriptstyle k})^2 
+\left( \fr{1}{l_1}\fr{d\epsilon}{dk}\right)^2\right]^{-1}
 \ \ ,
\ee
where $l_1$ is the Anderson localization length (\ref{l1}). This is
the {\it averaged} value of $|C(k)|^2$ and therefore it can not be
used directly for our following calculations. However, (\ref{gredesk})
gives the proper estimate of the amplitude of the wave function in the
momentum representation $C(k)$ in the whole range of variation of the
momentum $k$.

In order to illustrate the physical origin of (\ref{gredesk}) it is
useful to rewrite the single particle Schr\"odinger equation in the
plane wave basis:
\begin{equation}\label{core}
C(k) = \sum_{q} \fr{\langle k | \hat{w} | q
\rangle}{\epsilon-\epsilon_{\scriptscriptstyle k}} C(q) \ \ \ ,
\end{equation}
where by $\hat{w}$ we have denoted the random potential (\ref{H1}) and
$\epsilon$ is the single particle energy corresponding to this
eigenfunction. In general this equation is not easier to solve than
the original Schr\"odinger equation with the Anderson Hamiltonian
(\ref{H1}). However, in order to extract the information we need, it
is enough to observe, that the r.h.s. of (\ref{core}) is saturated by
the only $C(q)$ with $q-k_0 \sim 1/l_1$, and (for negative $q$) $q+k_0
\sim 1/l_1$ , where $k_0$ is the first positive solution of the
equation $\epsilon_0 = -2t \cos k_{0}$ . Therefore, far away from the
region $|k \pm k_0| \sim 1/l_1$ the amplitude of $C(k)$ is determined
by the energy denominator in (\ref{core}) in accordance with
(\ref{gredesk}).

Now it is easy to write down the estimate for the amplitudes $C(k)$.
\bq\label{Ck1}
C(k) \sim \sqrt{\fr{l_1}{L}} \ \ \ 
{\rm if} \ \ \ k \pm k_{\scriptscriptstyle 0} \sim 1/l_1 \ \ , 
\ee
\bq\label{Ck2}
C(k) \sim \sqrt{\fr{l_1}{L}} \fr{1}{(k_{\scriptscriptstyle 0}\pm
k) l_1} \ \ \ 
{\rm if} \ \ \  1/l_1 \ll |k{\scriptscriptstyle 0} \pm k| 
\ll 1 \ \ . 
\ee
Everywhere except the Section 5 we suppose that $L\gg l_1$. The
formula (\ref{Ck1}) follows directly from the normalization condition
$\sum|C(k)|^2=1$. The denominator $k_{\scriptscriptstyle 0}\pm k$ in
(\ref{Ck2}) stands for $\epsilon-\epsilon_{\scriptscriptstyle k}$ in
(\ref{core}) and all the other factors may be found by comparison of
(\ref{Ck2}) with (\ref{Ck1}) at $|k \pm k_{\scriptscriptstyle 0}| \sim
1/l_1 $. In general, for arbitrary $|k \pm k_{\scriptscriptstyle 0}|
\sim 1$
\bq\label{Ck3}
C(k) \sim \sqrt{\fr{1}{l_1 L}} \ \ \ .
\ee

Now let us take into account the interaction $V_{int}$ . It is natural
to use the basis of products of single particle localized states
\bq\label{pbasis}
|1,2\rangle = \psi_1(n_1) \psi_2(n_2) \ \ \ .
\ee
This wave function describes distinguishable particles. The
generalization to the case of bosons or electrons with opposite spins
is straightforward. Moreover, in the following section we will
effectively turn to the bosonic case because the Hubbard short-range
interaction (\ref{Vint}) separates the symmetric and antisymmetric
eigenstates and the Hamiltonian (\ref{Ham}) is invariant under
particle permutation ($n_1\rightarrow n_2 \ , \ n_2 \rightarrow n_1$).

In terms of the summation over the original lattice-sites the matrix
element between the states (\ref{pbasis}) takes the form
\bq\label{me0}
\langle 3,4| V_{int} |1,2 \rangle =
U \sum_n
\psi_1(n) \psi_2(n)\psi_3(n) \psi_4(n) \ \ .
\ee
Each wave function here decays exponentially outside the segment of
the lattice of  length $\sim l_1$. Therefore the matrix element
(\ref{me0}) vanishes if not all the functions $\psi_i(n)$
overlap. Everywhere in the following we consider only the matrix
elements for overlapping states.

In order to estimate the matrix element (\ref{me0}), the author of
\cite{Dima} assumed that each wave function $\psi_i(n)$ within its
segment $\sim l_1$ is completely random. Under this assumption one
immediately finds the following estimate
\bq\label{meDima}
\langle 3,4| V_{int} |1,2 \rangle_{chaotic} \sim 
\fr{U}{l_1^{3/2}} \ \ .
\ee
On the other hand, the assumption itself about chaoticity of the wave
function is evidently inconsistent with the accurate estimate of the
Fourier transform (\ref{gredesk}) - (\ref{Ck3}).

The difficulties in estimation of the matrix element (\ref{me0})
originate from the almost regular and fast oscillations of
the single particle wave functions. It may be seen immediately from
any toy example, that the matrix element for oscillating functions is
usually much more suppressed than for random ones. On the other hand,
for oscillating wave functions it is natural to consider the matrix
element (\ref{me0}) in terms of plane wave amplitudes (\ref{pwbasis})
\bq\label{me}
\langle 3,4| V_{int} |1,2 \rangle = \fr{U}{L} 
\sum_{k_i}
C_3^{\star}({k_{\scriptscriptstyle 3}})
C_4^{\star}({k_{\scriptscriptstyle 4}})
C_1({k_{\scriptscriptstyle 1}})
C_2({k_{\scriptscriptstyle 2}}) 
\delta_{k_1+k_2 ,k_3 +k_4} \ \
. 
\ee
Here the $\delta$-function accounts for the conservation of the total
momentum by the interaction $V_{int}$. One may use (\ref{n0}) in order
to demonstrate in terms of $C(k)$ that only the matrix elements
between spatially overlapping states $\psi_1,\ \psi_2,\ \psi_3,\
\psi_4$ survive.

Equation (\ref{me}) together with the plane wave amplitudes
(\ref{Ck1})-(\ref{Ck3}) is enough for the rough estimate of the matrix
element of $V_{int}$. As we have said, we will consider only the
overlapping states ($n_{0_i} - n_{0_j} \sim l_1$ (\ref{n0})),
otherwise the matrix elements decay exponentially.  We have considered
three different estimates (\ref{Ck1}), (\ref{Ck2}), and (\ref{Ck3}) of
the plane wave amplitude $C(k)$ for different ranges of variation of
the momentum $k$. Correspondingly, now we are going to present the
three different estimates of the matrix element (\ref{me}). First of
all, consider the single particle states
$|1\rangle,|2\rangle,|3\rangle,|4\rangle$ for which the momentum is
almost conserved in the transition (\ref{me}). This means that
$k_{\scriptscriptstyle 0_1}+k_{\scriptscriptstyle 0_2}\approx
k_{\scriptscriptstyle 0_3}+k_{\scriptscriptstyle 0_4}$ (the momentum
$k_{\scriptscriptstyle 0}$ is connected with the single particle
energy via (\ref {eps})). As we know each function $C_i(k)$ consists
of two narrow ($\Delta k \sim l_1^{-1}$) peaks. Just for
$k_{\scriptscriptstyle 0_1}+k_{\scriptscriptstyle 0_2}\approx
k_{\scriptscriptstyle 0_3}+k_{\scriptscriptstyle 0_4}$ all this peaks
overlap and therefore
\begin{eqnarray}\label{es1}
k_{\scriptscriptstyle 0_1}+k_{\scriptscriptstyle 0_2}
-k_{\scriptscriptstyle 0_3}-k_{\scriptscriptstyle 0_4} 
&\sim& 1/l_1 \ \ , \nonumber \\ 
\langle 34| V_{int} | 12 \rangle \sim 
\fr{U}{L} \left( \fr{l_1}{L}\right)^2 \left(\fr{L}{l_1}\right)^3
&\sim& \fr{U}{l_1} \ \ .
\end{eqnarray}
Here the first factor $U/L$ comes directly from (\ref{me}),
$(l_1/L)^2$ is the fourth power of the wave function at the maximum
(\ref{Ck1}), and $(L/l_1)^3$ is the effective number of terms in the
sum in (\ref{me}).  Suppose now that the momentum conservation is
completely violated $\Delta k \sim 1$.  In this case the peaks for all
$C(k)$ could not overlap simultaneously. One of the functions should
be taken at the tail (\ref{Ck3}):
\begin{eqnarray}\label{es3}
k_{\scriptscriptstyle 0_1}+k_{\scriptscriptstyle 0_2}
-k_{\scriptscriptstyle 0_3}-k_{\scriptscriptstyle 0_4} 
&\sim& 1 \ \ , \nonumber \\
\langle 34| V_{int} | 12 \rangle \sim 
\fr{U}{L} \left( \fr{l_1}{L}\right)^2 \fr{4}{l_1}
\left(\fr{L}{l_1}\right)^3
&\sim& \fr{U}{l_1^2} \ \ .
\end{eqnarray}
Here compared to (\ref{es1}) we have replaced one of the plane wave
amplitudes (\ref{Ck1}) by (\ref{Ck3}). The factor $4$ in (\ref{es3})
should not be considered very seriously, it simply symbolizes that
this replacement may be done in $4$ ways.

Finally, the most interesting case is if the momentum difference
$\Delta k = k_{\scriptscriptstyle 0_1}+k_{\scriptscriptstyle 0_2}-
k_{\scriptscriptstyle 0_3}-k_{\scriptscriptstyle 0_4}$ is much larger
than $1/l_1$ but still is small compared to $1$. To be more accurate
one should define a few $\Delta k$-s , $\Delta k = \pm
k_{\scriptscriptstyle 0_1}\pm k_{\scriptscriptstyle 0_2}\pm
k_{\scriptscriptstyle 0_3}\pm k_{\scriptscriptstyle 0_4}$. The matrix
element will be enhanced as in (\ref{es2}) below if at least one
$\Delta k$ is much smaller than $1$. We leave the consideration of
this complication until Section 4. The formula (\ref{Ck2}) for the
``short range tail'' now allows one to find the following estimate
\begin{eqnarray}\label{es2}
1/l_1 &\ll& |k_{\scriptscriptstyle 0_1}
+k_{\scriptscriptstyle 0_2}-k_{\scriptscriptstyle 0_3}
-k_{\scriptscriptstyle 0_4}| \ll 1 \ \ , \nonumber \\
\langle 3,4| V_{int} | 1,2 \rangle &\sim& 
\fr{U}{l_1^2} \fr{1}{k_{\scriptscriptstyle 0_1}+
k_{\scriptscriptstyle 0_2}-k_{\scriptscriptstyle 0_3}-
k_{\scriptscriptstyle 0_4}} \ \ .
\end{eqnarray}
Just this last matrix element will lead to the enhancement of the
two-particle localization length. We will return to the more accurate
and detailed estimate of the matrix element for the case (\ref{es2})
in  Section 4.

Note that all these estimates (\ref{es1}),(\ref{es3}),(\ref{es2})
differ from those of \cite{Dima} $\langle 3,4| V_{int} | 1,2
\rangle \sim U/l_1^{3/2}$ (\ref{meDima}).

The natural tool for the investigating the coherent propagation
proposed in \cite{Imry} is the Thouless block picture
\cite{Thouless}. Consider the two one-particle $1d$ Hamiltonians with
the interaction (\ref{Ham}) as the $2d$-Hamiltonian on a large
$L\times L$ square. Because the Hubbard interaction (\ref{Vint})
affects only the diagonal $n_1=n_2$ it is natural to use (for a
moment) the center of mass variables $\fr{1}{2} (n_1+n_2)$ and $(n_1
-n_2 )$. Now the large $2d$ system should be mentally divided into
square blocks of size $l_b\times l_b$ with $l_b\sim l_1$. Due to the
exponential decay of the localized eigenstates at $|n-n_0|>l_1$ and
the diagonal form of the interaction (\ref{Vint}), it is enough to
consider only one row of blocks with $|n_1-n_2|<l_b/2$. Thus our
$2d$-system reduces to quasi-$1d$ one. The $m$-th block is defined by
the inequalities
\begin{eqnarray}\label{block}
&\,& |n_1-n_2|<\fr{l_b}{2} \ \ , \\
&\,& ml_b<\fr{n_1+n_2}{2} < (m+1) l_b \ \ . \nonumber
\end{eqnarray}

About $l_b^2 \sim l_1^2$ states (\ref{pbasis}) fall down into each
block. Thus the typical level separation within one block is $\Delta
\sim t/l_1^2$. We suppose that the interaction (\ref{Vint}) mixes the
states from the same block and with about the same amplitude the
states from the nearest blocks. If one believes in the estimate
(\ref{meDima}) \cite{Dima} the delocalization (enhancement of the
coherent propagation length) follows immediately from the inequality
\cite{Imry}
\bq\label{ineqDima}
\langle 3,4| V_{int} |1,2 \rangle_{chaotic} \sim 
\Delta \sqrt{l_1} \gg \Delta \ \ .
\ee
However, unfortunately the matrix element (\ref{meDima}) is
inconsistent with the correct estimate.

Consider now, what one can conclude from the correct estimates
(\ref{es1})-(\ref{es2})? In order to understand whether the enhancement
exists or not, it is enough to fix one state (\ref{pbasis}) belonging,
say to the $m$-th block and to estimate the number of other states
mixed to that one with amplitude $\sim 1$. First of all, consider the
largest matrix element (\ref{es1}). This matrix element $\langle 34|
V_{int} |12 \rangle \sim 1/l_1$ is $l_1$ times larger than the typical
level spacing within one block. Nevertheless, the momentum
conservation necessary for (\ref{es1}) reduces in the same $l_1$ times
the number of states in one block available for this transition. Thus
\bq\label{row0}
\Delta_{eff_0} \sim \langle 3,4|
V_{int} |1,2 \rangle \sim \fr{1}{l_1} \ \ \ {\rm for} \ \ \ 
|k_{\scriptscriptstyle 0_1} +k_{\scriptscriptstyle 0_2} 
-k_{\scriptscriptstyle 0_3} -k_{\scriptscriptstyle 0_4}| 
<\fr{1}{l_1} \ \ ,
\ee
which means that about one state is strongly mixed to the given one by
this matrix element.

So we are looking for the admixture to the state $|1,2\rangle$
belonging to the $m$-th block.  Let us divide all the $\sim l_1^2$
states $|3,4\rangle$ from the $m$-th and two nearest blocks into
smaller portions in accordance with the momentum
non-conservation $\Delta k =k_{\scriptscriptstyle 0_1}
+k_{\scriptscriptstyle 0_2} -k_{\scriptscriptstyle 0_3}
-k_{\scriptscriptstyle 0_4}$. Into the $n$-th portion we will put the
states with $2^n/l_1 < |\Delta k| <2^{n+1}/l_1$. Due to (\ref{es2})
one may easily compare the effective level splitting $\Delta_{eff_n}$
for each portion with the corresponding matrix element
\begin{eqnarray}\label{rowall}
&\,&\Delta_{eff_1} \sim \langle 3,4|
V_{int} |1,2 \rangle \sim \fr{1}{2} \fr{1}{l_1} \ \ \ {\rm for} \ \ \ 
\fr{1}{l_1} <|k_{\scriptscriptstyle 0_1} +k_{\scriptscriptstyle 0_2} 
-k_{\scriptscriptstyle 0_3} -k_{\scriptscriptstyle 0_4}| 
<\fr{2}{l_1} \ \ , 
\nonumber \\
&\,&\Delta_{eff_2} \sim \langle 3,4|
V_{int} |1,2 \rangle \sim \fr{1}{4} \fr{1}{l_1} \ \ \ {\rm for} \ \ \ 
\fr{2}{l_1} <|k_{\scriptscriptstyle 0_1} +k_{\scriptscriptstyle 0_2} 
-k_{\scriptscriptstyle 0_3} -k_{\scriptscriptstyle 0_4}| 
<\fr{4}{l_1} \ \ ,
\\
&\,& \ \ \ \ \ \ \ \ \ \  .\, .\, .\, .\, .\, .\, .\, .\, .\, 
.\, .\, .\, .\, .\, .\, .\, .\, .\, .\, .\, .\, .   \nonumber \\
&\,& \ \ \ \ \ \ \ \ \ \  .\, .\, .\, .\, .\, .\, .\, .\, .\, 
.\, .\, .\, .\, .\, .\, .\, .\, .\, .\, .\, .\, .   \nonumber \\
&\,&\Delta_{eff_n} \sim \langle 3,4|
V_{int} |1,2\rangle \sim \fr{1}{2^{n}} \fr{1}{l_1} \ \ \ {\rm for} \ \ \
\fr{2^{n-1}}{l_1} <|k_{\scriptscriptstyle 0_1} 
+k_{\scriptscriptstyle 0_2} 
-k_{\scriptscriptstyle 0_3} -k_{\scriptscriptstyle 0_4}| 
<\fr{2^n}{l_1} \ \ .
\nonumber 
\end{eqnarray}
The number of states falling into the $n$-th portion increases like
$2^n$, but simultaneously the matrix element (\ref{es2}) decreases by
the same factor. The number of rows in the formula (\ref{rowall}) is
evidently $\ln(l_1)$ and therefore we see that each simple state
$|1,2\rangle$ (\ref{pbasis}) may be mixed with $\sim \ln(l_1) \gg 1$
others.

We will see in the Sections 5,6 how this $\sim \ln(l_1)$ enhancement
may lead to the coherent propagation length (\ref{lpr},\ref{l2new})

The approach of ref. \cite{Sushkov} is sometimes considered as one
which allows to get the better understanding of the TIP-problem. 
Therefore before going further we would like
to consider the validity of the method of \cite{Sushkov} in the light
of our estimate for the two-particle matrix element
(\ref{es1})-(\ref{es2}).

The authors ref.\cite{Sushkov} calculated
the averaged Breit-Wigner width $\Gamma$ for TIP in small $L\ll l_1$
sample. Connection with the full TIP problem is made via the relation
\bq\label{a}
l_2/l_1 \sim \Gamma \rho \ \ ,
\ee
where $\Gamma$ is the Breit-Wigner width for the sample of the size
$L\sim l_1$ and $\rho$ is the total density of two-particle states in
the same sample. The authors do not derive the eq. (\ref{a}), but find
the support for it in the
refs. \cite{Dima,Imry,Dima3,Fyodorov,Frahm}. The eq. (\ref{a}) seems
to be useful if the interaction would be able to mix with about the
same probability all overlapping two-particle states. However, as we
have shown in (\ref{es1},\ref{es3},\ref{es2}) the actual matrix
elements between different states are of the very different
magnitude. For example the total width $\Gamma$ in (\ref{a}) is
determined by a very small fraction $\sim 1/l_1$ of all two-particle
states. In its turn the main part of the density of states $\rho$ is
given by the states which may be mixed only by a very small matrix
element (\ref{es3}) (in $1/\sqrt{l_1}$ times smaller than is needed
for the estimate of ref. \cite{Dima}).

Thus, there seems to be no reason to join in one expression (\ref{a})
the $\Gamma$ and $\rho$ which have so different physical origin. One
may divide all the two-particle states into a classes so that the
matrix elements from the given state $|12\rangle$ to any other within
one class will be of the same order of magnitude. Just as we did in
the eq. (\ref{rowall}). Now the natural generalization of (\ref{a})
will be to introduce the large parameter (see (\ref{rowall}))
\bq\label{b}
g\sim \sum_{classes} \Gamma_i \rho_i \sim U/t \ln(l_1) \ \ ,
\ee
where $\Gamma_i$ is the partial width for the state $|12\rangle$ to
decay into the $i$-th class and $\rho_i $ is the density of states in
this class. This $g$ will work as the effective expansion parameter in
the perturbative treatment of the TIP problem. It is however still a
long way from eq. (\ref{b}) to the accurate estimate of the coherent
propagation length $l_2$. This will be just the main aim of our
Sections 5,6 to show that in order to find the TIP coherent
propagation length one should most naturally exponentiate $g$ and that
$l_2/l_1\sim \exp(const \times g)$ .

\section{The strong coupling limit}\label{sec:3}

The existing attempts to reduce the two particle problem to a random
matrix one were all based on the assumption that the mixing of the
simple states (\ref{pbasis}) due to the particle--particle interaction
is sufficiently random.

On the other hand, if there is no disorder ($w_n =0$) the exact
eigenfunctions of the Hamiltonian (\ref{Ham}) are easy to found in
terms of the variables $n_1-n_2$ and $n_1+n_2$ . Among these solutions
the eigenfunctions decaying exponentially at large $|n_1 -n_2|$ form
the molecular bound states sub-band. All remaining wave functions from
the continuous spectrum will be formed by rather regular combinations
of $\sin (n_{1,2})$ and $\cos (n_{1,2})$, though having a finite kink
at $n_1 = n_2$~. Thus for $w_n \equiv 0$ the particle--particle
interaction leads to considerable but rather regular rearrangement of
the noninteracting product basis set (\ref{pbasis}). It is evident
that for finite but small disorder ($|w_n| \ll t$) the regular
modification of the wave--functions should survive, at least in some
sense.

This regular rearrangement of the noninteracting basis may be thought
as the mixing of each state (\ref{pbasis}) with many others via the
largest matrix elements (\ref{es1}). The energy denominator for this
many states in general is not small (up to $\Delta E \sim t$) and thus
each individual admixture is added to (\ref{pbasis}) with a rather
small amplitude. However, due to the large number of these effectively
regular contributions, the final change of the wave function may be of
the order of one.

Thus we divide (although slightly arbitrarily) all the admixtures to
the given state $|1,2\rangle$ into two classes. First, those with
small energy denominator $\Delta E < 1/l_1$. These contribution are
completely random as we show in the following section and lead to the
large coherent propagation length. 

The second type of admixtures are those with large energy denominator
$\Delta E > 1/l_1$. We see no mechanism, how these corrections may
lead to the enhancement of the propagation length (although also we
have no rigorous proof that such mechanism does not exist). Moreover,
we are going to show in this section that these regular effects in the
strong coupling limit should lead to the suppression of the two
particle propagation. Up to now the authors of all the papers devoted
to the TIP problem have examined only the monotonous dependence of
$l_2$ on $U$ (say $l_2\sim U^2$ in \cite{Dima} and $l_2\sim |U|$ in
\cite{Oppen}). Among other papers the numerical result of
ref. \cite{Oppen} may be considered as the most serious objection to
our prediction (\ref{l2new}). However, our main statement in this
section is that all quantities characterizing the TIP problem should
depend in some complicated way on $U/t$ even for $U\sim t$. For
example $\gamma$ in (\ref{l2new}) for $U<t$ may be of the form (we do
not worry now about the sign of $U$)
\bq\label{gammaU}
\gamma(U)=\gamma_1\fr{U}{t}+\gamma_2\left(\fr{U}{t}\right)^2+
\gamma_3\left(\fr{U}{t}\right)^3+ ... \ \ ,
\ee
where all $\gamma_1,\gamma_2, \ ...$ are of the order of one $\gamma_i
\sim 1$. The same holds for the overall normalization of
(\ref{l2new}).  Therefore, because the authors of \cite{Oppen} have
not taken into account this complicated dependence of $l_2$ on $U$, we
believe that they have used the irrelevant fitting function for the
$l_1$ dependence as well (see also the discussion \cite{discussion}).

Let us consider now the case of very strong Hubbard interaction
(\ref{Vint}) $U\gg t$. Consider also the new basis set of states
relevant to this limiting case. First of all, the trivial subset of
diagonal states evidently decouples for the large $U$
\bq\label{diagonal}
| n \rangle = \delta_{n,n_1} \delta_{n,n_2} \ \ ,
\ee
where $n_1 , n_2$ are the lattice variables for two particles. These
states form a narrow molecular zone with  energy close to $U$ and
the effective molecule mass  close to $\sim U/t^2$. If $t^2/U
\ll w$ these particles do not move. If $t^2/U \gg w$,  Anderson
localization takes place, though the localization length is small
compared to (\ref{l1}).

The most surprising fact is that in the model under consideration all
the remaining eigenstates in the strong coupling limit may be found
exactly.  Let $\psi_i(n)$ be a set of localized single particle
states. Then it is easy to construct the set of antisymmetric states
\bq\label{anti}
|1,2\rangle^A = \fr{1}{\sqrt{2}}[\psi_1(n_1)\psi_2(n_2) -
\psi_2(n_1)\psi_1(n_2)]  \,\,\,\,\,\, ,
\ee
which are obviously the exact eigenfunctions of the Hamiltonian
(\ref{Ham}) even for arbitrary $U$. In addition to (\ref{diagonal})
and (\ref{anti}) let us consider the symmetric functions
\bq\label{symm}
|1,2 \rangle^S = \fr{1}{\sqrt{2}} sign(n_1-n_2) 
\big[ \psi_1(n_1)
\psi_2(n_2) - \psi_2(n_1) \psi_1(n_2) \big] \ \ ,
\ee
where $sign(x)=x/|x|$ and $sign(0)=0$.  First of all, the states
(\ref{diagonal})-(\ref{symm}) form the complete orthogonal set and are
all localized. Any two vectors from (\ref{diagonal}) and (\ref{anti})
are orthogonal by construction. The states (\ref{symm}) are evidently
orthogonal to (\ref{anti}) and (\ref{diagonal}). Finally, the integral
of the product of any two symmetric states (\ref{symm}) coincides with
the one for corresponding antisymmetric states. Therefore,
(\ref{diagonal})-(\ref{symm}) may be used as a new basis instead of
(\ref{pbasis}). Moreover, as we will show now the symmetric functions
(\ref{symm}) are also the exact eigenfunctions of (\ref{Ham}) for
$U\rightarrow \infty$ (or $U \gg t$).

The physical explanation, why (\ref{symm}) are the exact
eigenfunctions for large $U$ is almost trivial. Consider our two
particles in $1d$ as  one particle on a $2d$ square $L\times L$. For
infinite $U$ this particle simply could not penetrate through the
barrier (\ref{Vint}) along the diagonal of the square. As we said
before, the antisymmetric states (\ref{anti}) are by construction the
exact solutions. But now, because the two parts of the volume are
separated by an opaque barrier, the symmetric states (\ref{symm}),
as well as any linear combination of (\ref{anti}) and (\ref{symm}),
become  exact eigenfunctions of the Hamiltonian (\ref{Ham}).  

However, this simple explanation does not allow one to find the
corrections to (\ref{symm}) for large but finite $U$. Therefore below
we calculate the matrix element of the effective interaction between
the two states $| 1,2 \rangle^S$ and $| 3,4 \rangle^S$ for $U\gg
t$. Consider the total Hamiltonian $H_{tot}$ (\ref{Ham}) as a matrix
in the basis (\ref{diagonal})-(\ref{symm}). It is enough to consider
only the symmetric states (\ref{diagonal}) and (\ref{symm}). The only
surviving non-diagonal elements of this matrix are
\bq\label{duraki}
\langle n | H_{tot} | 1,2 \rangle^S =
-\sqrt{2} t \big[ (\psi_1(n+1) - \psi_1(n-1))\psi_2(n)-
(\psi_2(n+1) - \psi_2(n-1))\psi_1(n) \big] \ .
\ee
Besides that there are evidently only the diagonal elements 
\begin{eqnarray}\label{idioti}
\langle n | H_{tot} | n \rangle = U+w_n \approx U \ \ &,& \\
\langle 3,4 |^S \, H_{tot} | 1,2 \rangle^S = 
\epsilon_{\scriptscriptstyle 1}+\epsilon_{\scriptscriptstyle 2} 
\ \ &.& \nonumber
\end{eqnarray}
Thus the mixing of the two states $| 1,2 \rangle^S$ and $| 3,4
\rangle^S$ may appear only in the second order of perturbation theory.
Consider for simplicity the case $\epsilon_{\scriptscriptstyle
1}+\epsilon_{\scriptscriptstyle 2} \approx
\epsilon_{\scriptscriptstyle 3}+\epsilon_{\scriptscriptstyle 4}$ . The
simple perturbative calculation immediately leads to the effective
interaction
\begin{eqnarray}\label{mestrong}
&\,&\langle 3,4 |^S \, U_{eff} | 1,2 \rangle^S = 
 \sum_n \fr{\langle 3,4 |^S \, H_{tot} | n \rangle
\langle n | H_{tot} | 1,2 \rangle^S}{U-
\epsilon_{\scriptscriptstyle 1}-\epsilon_{\scriptscriptstyle 2}} 
 = \fr{2t^2}{U-
\epsilon_{\scriptscriptstyle 1}-\epsilon_{\scriptscriptstyle 2}} 
\fr{4}{L} \\ &\,& \times \sum_{k_1+k_2=k_3+k_4} 
C_1(k_{\scriptscriptstyle 1})C_2(k_{\scriptscriptstyle 2})
C_3^{\star}(k_{\scriptscriptstyle 3}) 
C_4^{\star}(k_{\scriptscriptstyle 4}) 
[\sin k_{\scriptscriptstyle 3} -
\sin k_{\scriptscriptstyle 4}]  
[\sin k_{\scriptscriptstyle 1} -
\sin k_{\scriptscriptstyle 2}] \ . \nonumber
\end{eqnarray}
This effective matrix element should be compared with
(\ref{me}). First of all we see that the strength of the interaction
$U$ has been replaced by $t^2/U$. Thus the corrections to (\ref{symm})
are of the order of $t/U\ll 1$. In particular the estimate of the
matrix element (\ref{es1}-\ref{es2}) evidently holds for
(\ref{mestrong}) up to the trivial replacement $U\rightarrow t^2/U$
(see also (\ref{U2str}) below).

Another important difference between (\ref{me}) and (\ref{mestrong})
are the $\sin(k_i)$-s, which can not lead to any enhancement and even
tends to suppress the matrix element for $\epsilon_{\scriptscriptstyle
1}\approx \epsilon_{\scriptscriptstyle 2}$ and
$\epsilon_{\scriptscriptstyle 3}\approx \epsilon_{\scriptscriptstyle
4}$ .

Thus we see that the problem of TIP should be treated within the two
sufficiently different (somehow dual) approaches in the weak
(\ref{me0},\ref{me}) and strong (\ref{mestrong}) interaction
cases. However the $l_1$ dependence of the matrix element estimated in
(\ref{es1}-\ref{es2}) coincides for both limits. 

In the intermediate region $U\sim t$ both formulas (\ref{me}) and
(\ref{mestrong}) should be modified. In order to illustrate the nature
of this modification consider for example the case $U<t$. Let us
consider also the eigenstates with the two-particle energy close to
some value $E_0$. It is convenient to consider the narrow strip
in energy of the width $2\delta_E \sim t/l_1$ around $E_0$. As we have
told at the beginning of this section the states within the strip
$|E_{ij}-E_0|<\delta_E$ will be mixed chaotically. The only role
played by the other states $|E_{ij}-E_0|>\delta_E$ is to renormalize
regularly the effective matrix element between chaotically mixed
ones. Let the states $| 1,2 \rangle$ and $| 3,4 \rangle$ belong to the
strip $|E_{12}-E_0|,|E_{34}-E_0|<\delta_E$ ($| 1,2 \rangle$ and $| 3,4
\rangle$ are the simple products of noninteracting localized single
particle states). Than one may write perturbatively the renormalized
interaction
\begin{eqnarray}\label{merenorm}
\langle 3,4 | \, V_{ren} | 1,2 \rangle &=&
\langle 3,4 | \, V_{int} | 1,2 \rangle  + \\
&\,& + \sum_{|E_{ij}-E_0|>\delta_E}
\fr{\langle 3,4 | \, V_{int} | i,j \rangle 
\langle i,j | \, V_{int} | 1,2 \rangle }{E_0-E_{ij}}+ ... \ \ .
\nonumber
\end{eqnarray}
The most interesting for us situation is when the momentum in the
transition $| 1,2 \rangle\rightarrow | 3,4 \rangle$ is not conserved
and the leading contribution in (\ref{merenorm}) is described by
(\ref{es2}). In this case one of the matrix elements in the sum in
(\ref{merenorm}) may be large like $U/l_1$ (\ref{es1}) (and the small
factor $1/l_1$ in it may be compensated due to a number of terms in
the sum). However, as one may easily see, the second matrix element in
the sum should again be of the form (\ref{es2}). Thus we see no
reason, why the renormalization of the bare matrix element
(\ref{me0},\ref{me},\ref{es2}) will not be described by the series of
the kind $1+(U/t)+(U/t)^2+ ...$ .

It is clear that the states (\ref{symm}) should be also the proper
solutions for the continuous version of the model (\ref{Ham}) with
strong repulsive interaction $V_{int}=G\delta(x-y)$. Of course in this
case the energy $\epsilon_{\scriptscriptstyle
1}+\epsilon_{\scriptscriptstyle 2}$ should be also sufficiently small
($\epsilon_{\scriptscriptstyle 1}+\epsilon_{\scriptscriptstyle 2}\ll
G^2$). In fact the continuous model may be simply considered as the
low energy limit of our lattice model (\ref{Ham}).

To conclude this section let us say again that the comparison of our
eq. (\ref{mestrong}) with (\ref{me}) shows that the crossover from
weak to strong coupling regime takes place somewhere at $U/t\sim
1$. The authors of refs. \cite{Oppen,Sushkov} have mentioned that they
would expect some deviations from their predictions at large
$U$. However as we can understand, they have in mind just the case
$|U|\gg t$. Moreover, one would even hardly found from
\cite{Oppen,Sushkov}, what is the sign of these expected corrections.

\section{Estimation of the matrix element}\label{sec:4}

The rough estimate of the matrix element (\ref{es2}) in principle is
enough to formulate the effective matrix model which will be
considered in the following two sections. Nevertheless, because the
distribution of the matrix elements is of crucial importance for our
result, we would like to present in this section a more accurate 
estimate than (\ref{es2}). We will consider both the matrix element of
the interaction potential (\ref{Vint}) in the basis of the simple
products of noninteracting single-particle states (\ref{pbasis}) and
the matrix element of the effective interaction in the strong coupling
case (\ref{mestrong}).

Although, our formulas (\ref{U2}),(\ref{U2nc}) below are too detailed
to be used for the mapping onto the matrix model (Sections 5,6), 
they may be a good base for the
comparison with the numerical experiment, say, of the kind mentioned
in \cite{Pichard}.

Our main problem is to make a reliable estimate of the matrix
element with oscillating wave functions.  As we have said in 
Section 2, we are mostly interested in the case when the four plane
wave amplitudes $C_i(k)$ do not overlap strongly in the matrix element
(\ref{me}). This means that at least one of the wave functions should
be taken at the tail (\ref{Ck2}). For this case the Schr\"odinger
equation (\ref{core}) may be further simplified
\begin{equation}\label{core1}
C(k)_{tail} = \sum_{|q-k_0|<\delta} \fr{\langle k | \hat{w} | q
\rangle}{\epsilon-\epsilon_{\scriptscriptstyle k}} C(q)_{core} +
\sum_{|q+k_0|<\delta} \fr{\langle k | \hat{w} | q
\rangle}{\epsilon-\epsilon_{\scriptscriptstyle k}} C(q)_{core} \ \ \ ,
\end{equation}
where $|k\pm k_{\scriptscriptstyle 0}| \gg 1/l_1$ and $\delta \sim
1/l_1$ (say $\delta = 5/l_1$ ). Here we have divided each wave
function $C(k)$ into two parts. One part is ``the core'', or the
largest part of the wave function coming from
$|k-k_{\scriptscriptstyle 0}|<\delta$ and $|k+k_{\scriptscriptstyle
0}|<\delta$. Another part is the tail at $|k\pm k_{\scriptscriptstyle
0}|>\delta$~.

The matrix element of the random potential between the two plane
waves which appeared in (\ref{core1}) is equal to
\begin{equation}\label{Wk}
\langle k^{\prime}|\hat w|k \rangle \equiv
\langle 0|\hat w|k-k^{\prime} \rangle  ={1\over{L}}\sum_{n=1}^L w_n
e^{i(k-k^{\prime})n}.
\end{equation}
For the quantitative calculation of the matrix element of the
interaction potential it is important that (\ref{Wk}) depends only on
the one variable $k-k^{\prime}$. The distribution of $\langle
k^{\prime}|\hat w|k \rangle$ is an usual normal one with variance (see
(\ref{weak}))
\begin{equation}\label{w2}
\overline {|\langle k^{\prime}|\hat w|k \rangle|^2}=w^2/L.
\end{equation}

Formula (\ref{core1}) allows us to demonstrate one important
difference between the ``core'' and the ``tail'' of single particle
wave function. The main part of the wave function is almost regular
(one should not worry about the trivial randomness due to the position
of the center of the wave packet $n_0$ (\ref{n0}) because we consider
only overlapping functions). On the other hand the tail (\ref{core1})
due to $\langle 0|\hat w|k-k^{\prime} \rangle $ is sufficiently
random. More accurately, $C(k)$ and $C(k^{\prime})$ are correlated
only for $k-k^{\prime}\sim 1/l_1$ and are not correlated for
$|k-k^{\prime}| \gg 1/l_1$.

This observation allows us to define a relatively simple procedure of
averaging the squared matrix element (\ref{me0}),(\ref{me}). The
mathematically rigorous way of averaging the $|\langle
3,4|V_{int}|1,2\rangle|^2$ would be to introduce eight collective
variables~: centers of the wave packets $n_{\scriptscriptstyle
0_1},n_{\scriptscriptstyle 0_2},n_{\scriptscriptstyle
0_3},n_{\scriptscriptstyle 0_4}$ and the energies
$\epsilon_{\scriptscriptstyle 1},\epsilon_{\scriptscriptstyle
2},\epsilon_{\scriptscriptstyle 3},\epsilon_{\scriptscriptstyle
4}$. The ``rigorous'' averaging should then be performed over the
ensemble of random Hamiltonians (\ref{H1}) for fixed values of these
collective variables. However, the center of the packet and the energy
are almost completely determined by the smooth core of the wave
function. The actual randomness of the matrix element, as we will see
below, is hidden in the weak tail (\ref{core1}) which almost does not
affect $n_0$ and $\epsilon$. Therefore in this section we are going to
consider the distribution of the matrix elements $\langle
3,4|V_{int}|1,2\rangle$ for ``frozen'' cores. In terms of the single
particle matrix elements (\ref{Wk}), it means that we have frozen the
values of $\langle 0|\hat w| \Delta \rangle $ in a few small ($\sim
1/l_1$) segments around $\Delta = 0$ and $\Delta = \pm
2k_{\scriptscriptstyle 0_{i}}\ (i=1-4)$ and have performed the
averaging over realizations of all the others elements $\langle 0|\hat
w| \Delta' \rangle $.

As in the previous sections, consider the transition between the
states $|1,2\rangle$ and $|3,4\rangle$. Let now $|1,2\rangle$ and
$|3,4\rangle$ be the symmetrized simple products of the kind
(\ref{pbasis}):
\bq\label{spbasis}
|1,2\rangle =\fr{1}{\sqrt{2}}\big( 
\psi_1(n_{\scriptscriptstyle 1}) 
\psi_2(n_{\scriptscriptstyle 2})+
\psi_2(n_{\scriptscriptstyle 1}) 
\psi_1(n_{\scriptscriptstyle 2})\big) \ \ \ .
\ee
This symmetrization  simply adds an overall factor $2$ to the
matrix element (\ref{me0},\ref{me}).

We are interested in the matrix elements between the approximate
eigenstates which may be mixed strongly by the interaction. Therefore,
because the largest matrix element (\ref{es1}) is of the order of
$\sim 1/l_1$, the product states should be almost degenerate
$\epsilon_{\scriptscriptstyle 1}+ \epsilon_{\scriptscriptstyle 2} -
\epsilon_{\scriptscriptstyle 3} -\epsilon_{\scriptscriptstyle 4} \sim
t/l_1$. Moreover, because we have seen in Section 2 that the smallest
matrix elements (\ref{es3}) themselves could not lead to any
considerable enhancement, it is enough to choose
$|\epsilon_{\scriptscriptstyle 1} - \epsilon_{\scriptscriptstyle 3}|
\ll t$ and $|\epsilon_{\scriptscriptstyle 2} -
\epsilon_{\scriptscriptstyle 4}| \ll t$ (although still $t/l_1 \ll
|\epsilon_{\scriptscriptstyle 1} - \epsilon_{\scriptscriptstyle
3}|,|\epsilon_{\scriptscriptstyle 2} - \epsilon_{\scriptscriptstyle
4}|$). 

 We have divided each wave function in (\ref{core1}) into two parts: a
strong ``core'' and a weak ($\sim w/\Delta\epsilon$) ``tail''. Due to
momentum conservation in the matrix element (\ref{me}), direct
transition between cores is not allowed now. Equation (\ref{core1})
shows however that this transition effectively may take place in the
first order of perturbation theory in $\langle k | \hat{w} | q
\rangle$. Substitution of (\ref{core1}) into (\ref{me}) (with the
additional factor of $2$ due to the symmetrization (\ref{spbasis}))
gives
\begin{eqnarray}\label{iii}
\langle 3,4|V_{int}|1,2\rangle = 2
{U\over L}\sum_{\Delta} \sum_{q_i=core}
C_3^{\star}(q_{\scriptscriptstyle 3}) 
C_4^{\star}(q_{\scriptscriptstyle 4})
C_1(q_{\scriptscriptstyle 1})
C_2(q_{\scriptscriptstyle 2})
 \delta_{\Delta,q_3+q_4-q_1-q_2}\\
\times \langle 0 | \hat w | \Delta \rangle \left\{
{{1}\over{\epsilon_{\scriptscriptstyle 1}-
\epsilon_{\scriptscriptstyle q_1+\Delta}}} +
{{1}\over{\epsilon_{\scriptscriptstyle 3}-
\epsilon_{\scriptscriptstyle q_3-\Delta}}} +
{{1}\over{\epsilon_{\scriptscriptstyle 2}-
\epsilon_{\scriptscriptstyle q_2+\Delta}}} +
{{1}\over{\epsilon_{\scriptscriptstyle 4}
-\epsilon_{\scriptscriptstyle q_4-\Delta}}} \right\} \ \ .
\nonumber  
\end{eqnarray}
Here the summation over $q_i$ includes only the largest plane wave
amplitudes (\ref{Ck1}) $C(q)\sim \sqrt{l_1/L}$. Each of the four terms
in the figure brackets in (\ref{iii}) corresponds to the replacement
of one of the $C(q)$ in (\ref{me}) by the perturbative formula
(\ref{core1}). However, due to (\ref{Wk}) all these four contributions
have the same random weight $\langle 0 | \hat{w} | \Delta \rangle$.

In Section 2 we have classified all the matrix elements according to
the value of momentum non-conservation in the transition $\Delta k$. In
order to make the simplest estimates up to now we made almost no
difference between the momentum $k$ and the single particle energy
$\epsilon =-2t\cos(k)$. However, the matrix element (\ref{es2}) turns
out to be enhanced just due to the small energy denominator. Each wave
function in momentum representation consists of two peaks $k\approx
k_0$ and $k\approx - k_{\scriptscriptstyle 0}$ (we choose
$k_{\scriptscriptstyle 0}$ to be positive). Thus, effectively in the
equation (\ref{iii}), in addition to the four terms in the figure
brackets, we have $4\times 4 = 16$ transitions
\begin{eqnarray}\label{transition}
\left( \begin{array}{ll}
 \ \ k_{\scriptscriptstyle 0_1},\ \ k_{\scriptscriptstyle 0_2}\\
    -k_{\scriptscriptstyle 0_1},\ \ k_{\scriptscriptstyle 0_2} \\
 \ \ k_{\scriptscriptstyle 0_1},-k_{\scriptscriptstyle 0_2} \\
    -k_{\scriptscriptstyle 0_1},-k_{\scriptscriptstyle 0_2}               
               \end{array} \right) 
\rightarrow
\left( \begin{array}{ll}
 \ \ k_{\scriptscriptstyle 0_3},\ \ k_{\scriptscriptstyle 0_4} \\
    -k_{\scriptscriptstyle 0_3},\ \ k_{\scriptscriptstyle 0_4} \\
 \ \ k_{\scriptscriptstyle 0_3},-k_{\scriptscriptstyle 0_4} \\
    -k_{\scriptscriptstyle 0_3},-k_{\scriptscriptstyle 0_4}               
               \end{array} \right) 
\end{eqnarray}
between pairs of narrow peaks (cores). The transitions between these
pairs became possible due to perturbation theory (\ref{core1}), but
not all of them (although still many) will be enhanced by the small
energy denominators. On the other hand, the perturbative formula
(\ref{core1}) also contains the random matrix element $\langle 0 |
\hat{w} | \Delta \rangle$. This $\Delta$ will be different for
different transitions from (\ref{transition}) and thus different
contributions will not interfere in the averaged value of the squared
matrix element of $V_{int}$. This means that one may calculate
separately the contribution of each two-by-two pair in
(\ref{transition}) into $\overline{|\langle 3,4|
V_{int}|1,2\rangle|^2}$.

Consider first the case $(k_{\scriptscriptstyle
0_1},k_{\scriptscriptstyle 0_2}\rightarrow k_{\scriptscriptstyle
0_3},k_{\scriptscriptstyle 0_4})$ (only this case was supposed to be
considered in the rough estimate (\ref{es2}) of Section 2). All the
energy denominators in (\ref{iii}) now are small compared to the
typical energy $\sim t$. For example, this means that
\bq\label{v1}
\epsilon_{\scriptscriptstyle 1}-
\epsilon_{\scriptscriptstyle q_1+\Delta} \approx 
(k_{\scriptscriptstyle 0_1}-
q_{\scriptscriptstyle 1}-\Delta) v_{\scriptscriptstyle 1} 
\approx
-\Delta v_{\scriptscriptstyle 1} \approx
-(k_{\scriptscriptstyle 0_3}+k_{\scriptscriptstyle 0_4}-
k_{\scriptscriptstyle 0_1}-k_{\scriptscriptstyle 0_2}) 
v_{\scriptscriptstyle 1} \ \ \ ,
\ee
where we have introduced the velocity (see (\ref{eps}))
\bq\label{velocity}
v = \fr{d\epsilon}{dk} = 2t \sin k \ \ .
\ee
Substitution of (\ref{v1}) into (\ref{iii}) for the transition
$(k_{\scriptscriptstyle 0_1},k_{\scriptscriptstyle 0_2} \rightarrow
k_{\scriptscriptstyle 0_3},k_{\scriptscriptstyle 0_4})$ gives
\begin{eqnarray}\label{iiiv}
\langle 3,4|V_{int}|1,2\rangle _{\scriptscriptstyle ++,++}&=& 2
{U\over L}\sum_{\Delta} \sum_{|k_{0_i}-q_i|<\delta}
C_3^{\star}(q_{\scriptscriptstyle 3}) 
C_4^{\star}(q_{\scriptscriptstyle 4})
C_1(q_{\scriptscriptstyle 1})
C_2(q_{\scriptscriptstyle 2})
 \delta_{\Delta,q_3+q_4-q_1-q_2}\nonumber \\ &\,&
\fr{\langle 0 | \hat w | \Delta 
\rangle}{k_{\scriptscriptstyle 0_1}+
k_{\scriptscriptstyle 0_2}-
k_{\scriptscriptstyle 0_3}-
k_{\scriptscriptstyle 0_4}} \left\{
{{1}\over{v_{\scriptscriptstyle 1}}} -
{{1}\over{v_{\scriptscriptstyle 3}}} +
{{1}\over{v_{\scriptscriptstyle 2}}} -
{{1}\over{v_{\scriptscriptstyle 4}}} \right\} \ \ .
\end{eqnarray}
Here $(k_{\scriptscriptstyle 0_1}+k_{\scriptscriptstyle
0_2}-k_{\scriptscriptstyle 0_3}-k_{\scriptscriptstyle 0_4})^{-1}$ is
the large factor considered in eq. (\ref{es2}) of Section 2. The
subscript ${\scriptstyle ++,++}$ stands for the signs of
$(k_{\scriptscriptstyle 0_1},k_{\scriptscriptstyle 0_2})$ and $
(k_{\scriptscriptstyle 0_3},k_{\scriptscriptstyle 0_4})$ in the
transition.

However, if $k_{\scriptscriptstyle 0_1}+k_{\scriptscriptstyle
0_2}-k_{\scriptscriptstyle 0_3}-k_{\scriptscriptstyle 0_4} \ll 1$ and
$\epsilon_{\scriptscriptstyle 1}-\epsilon_{\scriptscriptstyle 3}
\approx \epsilon_{\scriptscriptstyle 4}-\epsilon_{\scriptscriptstyle
2} \ll t$ one has
\bq\label{vc}
v_{\scriptscriptstyle 1}-
v_{\scriptscriptstyle 3} \approx 
v_{\scriptscriptstyle 4}-
v_{\scriptscriptstyle 2} \sim  
k_{\scriptscriptstyle 0_1}+
k_{\scriptscriptstyle 0_2}-
k_{\scriptscriptstyle 0_3}-
k_{\scriptscriptstyle 0_4} 
\ \ \ .
\ee
Thus the small denominator in (\ref{iiiv}) turns out to be compensated
by the strong cancelation of the terms in the figure brackets. On the
other hand, this cancelation is specific only to the transitions
$(k_{\scriptscriptstyle 0_1},k_{\scriptscriptstyle 0_2} \rightarrow
k_{\scriptscriptstyle 0_3},k_{\scriptscriptstyle 0_4})$ and
$(-k_{\scriptscriptstyle 0_1},-k_{\scriptscriptstyle 0_2} \rightarrow
-k_{\scriptscriptstyle 0_3},-k_{\scriptscriptstyle 0_4})$ (or
$({\scriptstyle ++,++})$ and $({\scriptstyle --,--})$). For example,
let us consider the case $(k_{\scriptscriptstyle
0_1},k_{\scriptscriptstyle 0_2})\rightarrow (k_{\scriptscriptstyle
0_3},-k_{\scriptscriptstyle 0_4})$. Now $\Delta $ in (\ref{iii}) is of
the form
\bq\label{Delta}
\Delta_{\scriptscriptstyle ++,+-}= 
k_{\scriptscriptstyle 0_3}-
k_{\scriptscriptstyle 0_4}-
k_{\scriptscriptstyle 0_1}-
k_{\scriptscriptstyle 0_2}
\ \ \ .
\ee
Here $\Delta_{\scriptscriptstyle ++,+-}$ is not small: $\Delta \sim
1$. However, for $\Delta_{\scriptscriptstyle ++,+-}$ two of the energy
denominators in (\ref{iii}) are of the order of $ t$ but other two are
still small:
\begin{eqnarray}\label{pppm}
&\,&\epsilon_{\scriptscriptstyle 2}-
\epsilon_{ q_2+\Delta} \approx
\epsilon_{ -k_{0_2}}-
\epsilon_{ k_{0_2}+\Delta_{++,+-}} 
\approx (k_{\scriptscriptstyle 0_2}-
k_{\scriptscriptstyle 0_4}+
k_{\scriptscriptstyle 0_3}+
k_{\scriptscriptstyle 0_1}) v_2 \ \ , \\
&\,&\epsilon_{\scriptscriptstyle 4}-
\epsilon_{ q_4 -\Delta} \approx
\epsilon_{ k_{0_4}}-
\epsilon_{ -k_{0_4}-\Delta_{++,+-}} \approx
(k_{\scriptscriptstyle 0_4}-
k_{\scriptscriptstyle 0_2}+
k_{\scriptscriptstyle 0_3}-
k_{\scriptscriptstyle 0_1}) v_4 \ \ .   \nonumber
\end{eqnarray}
Simple substitution of (\ref{pppm}) into (\ref{iii}) gives
\begin{eqnarray}\label{iiivv}
\langle 3,4|V_{int}|1,2\rangle _{\scriptscriptstyle ++,+-}= 2
{U\over L}\sum_{\Delta} \ \sum_{q_{1,2,3} \approx k_{0_{1,2,3}}} \
\sum_{q_4 \approx -k_{0_4}} \nonumber \\ \times
C_3^{\star}(q_{\scriptscriptstyle 3}) 
C_4^{\star}(q_{\scriptscriptstyle 4})
C_1(q_{\scriptscriptstyle 1})
C_2(q_{\scriptscriptstyle 2})
 \delta_{\Delta,q_3+q_4-q_1-q_2}\ \ \ \ \ \\ \times
\fr{2(k_{\scriptscriptstyle 0_1}-
k_{\scriptscriptstyle 0_3})\langle 0 | \hat w | \Delta 
\rangle}{v_{\scriptscriptstyle 2}
[(k_{\scriptscriptstyle 0_1}-
k_{\scriptscriptstyle 0_3})^2-
(k_{\scriptscriptstyle 0_2}-
k_{\scriptscriptstyle 0_4})^2]} \ \ ,
\ \ \ \ \ \ \ \ \ 
\nonumber  
\end{eqnarray}
where we have made use of $v_{\scriptscriptstyle 2} \approx
v_{\scriptscriptstyle 4}$. Now it is convenient to go back from
momentum to coordinate space.  Equations (\ref{pwbasis}),(\ref{Wk})
allow one to rewrite (\ref{iiivv}) in the form
\begin{eqnarray}\label{iiicc}
&\,&\langle 3,4|V_{int}|1,2\rangle _{\scriptscriptstyle ++,+-}=  \\
&\,&{4U\over (\epsilon_{\scriptscriptstyle 1} -
\epsilon_{\scriptscriptstyle 3})}
\fr{v_{\scriptscriptstyle 1}v_{\scriptscriptstyle 2}}
{[v_{\scriptscriptstyle 1}^2-v_{\scriptscriptstyle 2}^2]}
\sum_{n} w_n  \psi^+_1(n) \psi^+_2(n)  
\psi^-_3(n) \psi^+_4(n) 
\ \ .
\nonumber  
\end{eqnarray}
Here we have used: $\epsilon_{\scriptscriptstyle 1}
-\epsilon_{\scriptscriptstyle 3} \approx \epsilon_{\scriptscriptstyle
2} -\epsilon_{\scriptscriptstyle 4}$ and, e.g.,
$\epsilon_{\scriptscriptstyle 1} -\epsilon_{\scriptscriptstyle
3}\approx v_1 (k_{\scriptscriptstyle 0_1} -k_{\scriptscriptstyle
0_3})$.  $\psi^+$ and $\psi^-$ are the positive and negative frequency
part of the single particle wave function respectively
($\psi^+=\psi^{-\star}$).

As we have mentioned before, due to the random factor $\langle 0 |
\hat w | \Delta \rangle$, the contribution (\ref{iiivv}) does not
interfere with others in the squared matrix element. On the other
hand, due to the random factor $w_n$ in (\ref{iiicc}), there are no
problems now with cancelation of the sum of regular oscillating
terms. Thus, $\psi^{\pm}$ in (\ref{iiicc}) are the almost regular
solutions of the single particle Schr\"odinger equation (``cores'')
which as we have seen are responsible mainly for the global features
of the eigen-vectors such as, e.g., the energy and the position of the
wave packet. Therefore one should naturally average the squared matrix
element via
\begin{eqnarray}\label{perduha}
\overline{|\sum_{n} w_n  \psi^+_1(n) \psi^+_2(n)  
\psi^-_3(n) \psi^+_4(n)|^2}=\\
w^2 \fr{1}{16}\sum_n \psi_1(n)^2 \psi_2(n)^2 \psi_3(n)^2 \psi_4(n)^2 
\ \ .\nonumber
\end{eqnarray}

We have considered in (\ref{iiicc}),(\ref{perduha}) only the
contribution of the transition with signature $({\scriptstyle
++,+-})$. It is easy to show that  identical contributions to the
variance of the matrix element come from the transitions
$({\scriptstyle --,-+})$, $({\scriptstyle +-,++})$, $({\scriptstyle
+-,--})$, $({\scriptstyle ++,-+})$, $({\scriptstyle --,+-})$,
$({\scriptstyle -+,++})$, and $({\scriptstyle +-,--})$. Finally, all
these eight contributions give
\bq\label{U2}
\overline{|\langle 3,4| V_{int}|1,2\rangle|^2} =
\fr{8 U^2 w^2}{(\epsilon_{\scriptscriptstyle 1} -
\epsilon_{\scriptscriptstyle 3})^2} 
\fr{v_{\scriptscriptstyle 1}^2 v_{\scriptscriptstyle 2}^2}
{[v_{\scriptscriptstyle 1}^2-v_{\scriptscriptstyle 2}^2]^2} 
\sum_n \psi_1(n)^2 \psi_2(n)^2 \psi_3(n)^2 \psi_4(n)^2
\ee
As we have said before the averaging here (the intermediate averaging)
is performed over the Fourier components of the random potential
$\langle 0 | \hat w | \Delta \rangle$, which do not contribute
sufficiently to the ``core'' part of the wave function, namely
$l_1|\Delta|\gg 1$ and $l_1|\Delta \pm 2 k_{\scriptscriptstyle
0_i}|\gg 1$.  The equation (\ref{U2}) may be used if $t/l_1 \ll
\epsilon_{\scriptscriptstyle 1} - \epsilon_{\scriptscriptstyle 3} \ll
t$ but the energy non-conservation in the transition is very small
$\epsilon_{\scriptscriptstyle 1}+\epsilon_{\scriptscriptstyle
2}-\epsilon_{\scriptscriptstyle 3}-\epsilon_{\scriptscriptstyle 4}
\sim t/l_1$. In particular one should not use this equation for
calculation of the Breit-Wigner width $\Gamma$ considered in
\cite{Sushkov}. As it was said at the end of Section 2 the total width
$\Gamma$ is determined by only the few largest matrix elements. In our
notations these are the matrix elements with
$\epsilon_{\scriptscriptstyle 1} - \epsilon_{\scriptscriptstyle 3}
\sim t/l_1$ and (\ref{U2}) may be considered in this case at best as
only the order of magnitude estimate.

One may easily repeat the calculation
(\ref{iiivv})-(\ref{U2}) for the less restrictive case
$\epsilon_{\scriptscriptstyle 1} - \epsilon_{\scriptscriptstyle 3}
\sim \epsilon_{\scriptscriptstyle 4} - \epsilon_{\scriptscriptstyle 2}
\sim \epsilon_{\scriptscriptstyle 1}+\epsilon_{\scriptscriptstyle
2}-\epsilon_{\scriptscriptstyle 3}-\epsilon_{\scriptscriptstyle 4} \ll
t$.
\begin{eqnarray}\label{U2nc}
&\,& \overline{|\langle 3,4| V_{int}|1,2\rangle|^2} = \\
&\,& 4  U^2 w^2 \fr{(\epsilon_{\scriptscriptstyle 1} -
\epsilon_{\scriptscriptstyle 3})^2+
(\epsilon_{\scriptscriptstyle 2}
-\epsilon_{\scriptscriptstyle 4})^2}{ 
v_{\scriptscriptstyle 1}^2 v_{\scriptscriptstyle 2}^2 \left( 
(\epsilon_{\scriptscriptstyle 1} -
\epsilon_{\scriptscriptstyle 3})^2/v_{\scriptscriptstyle 1}^2 - 
(\epsilon_{\scriptscriptstyle 2} -
\epsilon_{\scriptscriptstyle 4})^2/v_{\scriptscriptstyle 2}^2
\right)^2}  
\sum_n \psi_1(n)^2 \psi_2(n)^2 \psi_3(n)^2 \psi_4(n)^2 
\nonumber
\end{eqnarray}
In the following two sections we are going to consider the mappings of
the two-particle problem onto some special random matrix
models. However, these matrix models, although are expected to
reproduce qualitatively the main features of the two particle
eigenstates, certainly could not be used for any quantitative
calculation. Therefore it seems rather probable that one still should
perform in the future the accurate calculation of $l_2$ with the original
Hamiltonian (\ref{Ham}). The numerical verification of our expression
(\ref{U2nc}) seems to be a good starting point for such calculation.

In a similar way one may find the variance of the matrix elements
of the effective interaction (\ref{mestrong}) in the strong coupling
basis (\ref{symm}). We skip the details of this calculation and give
only the result
\bq\label{U2str}
\overline{|\langle 3,4|^S V_{int}|1,2\rangle^S|^2} =
\fr{4t^4 w^2}{U^2(\epsilon_{\scriptscriptstyle 1} -
\epsilon_{\scriptscriptstyle 3})^2} 
\fr{v_{\scriptscriptstyle 1}^2 v_{\scriptscriptstyle 2}^2}{t^4} 
\sum_n \psi_1(n)^2 \psi_2(n)^2 \psi_3(n)^2 \psi_4(n)^2 \ \ \ .
\ee
Here we have neglected $\epsilon_{\scriptscriptstyle
1}+\epsilon_{\scriptscriptstyle 2}$, compared to $U$ in the
denominator.

The matrix element in the simple basis of noninteracting states
(\ref{iiivv}),(\ref{U2}) is proportional to $(v_{\scriptscriptstyle
1}-v_{\scriptscriptstyle 2})^{-1}$. At first glance this factor may
effectively strengthen the interaction for
$\epsilon_{\scriptscriptstyle 1}\approx \epsilon_{\scriptscriptstyle
2}$. However, in the strong coupling limit (\ref{U2str}) all
denominators $(v_{\scriptscriptstyle 1}-v_{\scriptscriptstyle
2})^{-1}$ disappear. The reason for this more regular behavior of the
matrix element in the strong coupling limit is obvious. If
$v_{\scriptscriptstyle 1}$ is close to $v_{\scriptscriptstyle 2}$, the
relative kinetic energy of the two particles is small compared to the
interaction (\ref{Vint}) which is effectively equivalent to the strong
coupling case. In Section 3 we have introduced the new basis set
(\ref{diagonal}-\ref{symm}), which for $U\gg t$ is relevant for all
two-particle states. On the other hand, the factor
$(v_{\scriptscriptstyle 1}-v_{\scriptscriptstyle 2})^{-1}$ in
(\ref{iiivv}),(\ref{U2}) indicates that even for $t\gg U$ but 
$\epsilon_{\scriptscriptstyle 1}\approx \epsilon_{\scriptscriptstyle
2}$ one should modify the product basis (\ref{spbasis}). This 
modification consists of the strong mixing of many states
(\ref{spbasis}) with $\epsilon_{\scriptscriptstyle 1}\approx
\epsilon_{\scriptscriptstyle 2}$ in order to get something like
(\ref{symm}). However, we do not see how this mixing (rather regular
in fact) may lead to the enhancement of the coherent propagation
length.

The author of \cite{Imry}, in addition to the particle-particle propagation
has proposed to consider the coherent propagation of the
particle-hole pair. For a particle and a hole with excitation energy small
compared to $E_F$ one naturally has $v_{\scriptscriptstyle 1}\approx
v_{\scriptscriptstyle 2}$. However our strong coupling basis can not
be directly applied to this case because of the wave functions
(\ref{diagonal})-(\ref{symm}) do not describe the simple particle-hole
excitation above some vacuum. Thus the coherent propagation of particle
and hole may be very different from the coherent propagation of two
particles.

\section{Effective matrix model. Short sample}\label{sec:5}

The block picture considered at the end of Section 2 allows one to
perform easily the mapping of the two particle Hamiltonian (\ref{Ham})
onto the corresponding random matrix model. However, almost no results
may be obtained analytically for this matrix model as we will see in
Section 6. Therefore, we still would like to consider a simplified
version of the problem of coherent propagation. Namely, let two
interacting particles move inside the sufficiently short sample $L
\sim l_1$. In this case all particles are free to move throughout the
whole sample and there is no room for the interaction induced
delocalization. However, one may be interested in the chaotic mixing
of many simple states (like (\ref{pbasis}) or (\ref{symm})) with
different single particle energies $\epsilon$. The natural quantity,
characterizing the complexity of such states, is the so called inverse
participation ratio -- $l_{ipr}$ .  The explicit definition of
$l_{ipr}$ reads:
\bq\label{lipr} 
l_{ipr}^{-1} = \overline{\sum
|\alpha_n|^4} \ \ \ , 
\ee 
where $\alpha_n$ are the normalized coefficients of the expansion of
the chaotic eigenstate over the noninteracting two-particle states.

In order to find $l_{ipr}$ we are going to replace the solution of the
exact Schr\"odinger equation (\ref{Ham}) by the relevant random matrix
problem. First of all, as we have seen before (\ref{es1}) , the matrix
elements of $V_{int}$ never exceed the value $\langle 3,4| V_{int} |
1,2 \rangle \sim U/l_1$ (more concretely: $\max \langle 3,4| V_{eff} |
1,2 \rangle \sim \min(U,t^2/U)/l_1$). Therefore we may restrict our
attention to the consideration of only the part of the simple states,
having their total energy within the narrow band
\bq\label{delta} 
E_0 -\delta_E
< \epsilon_{\scriptscriptstyle 1} + 
\epsilon_{\scriptscriptstyle 2} < E_0 +\delta_E \ \ , 
\ee 
where $\delta_E \sim t/l_1$ . The total number of such states now is
$N \sim l_1$~. The matrix elements of the interaction potential
(\ref{me0}) or (\ref{mestrong}) between these states form the random
matrix. The main result of our previous consideration
(\ref{es2}),(\ref{U2}),(\ref{U2str}), is that the statistics of the
elements of this matrix is not uniform. In order to demonstrate
explicitly the structure of the interaction matrix,   one should
simply  enumerate the states (\ref{delta}) by assigning each one
a number $n$ in accordance with, say, the energy of the first
particle: $n(\epsilon_{\scriptscriptstyle 1}) >
n'(\epsilon_{\scriptscriptstyle 1}')$ if $\epsilon_{\scriptscriptstyle
1} > \epsilon_{\scriptscriptstyle 1}'$. Now, due to
(\ref{es2}),(\ref{U2}),(\ref{U2str}), the non-diagonal elements of our
matrix $M$ will be Gaussian distributed with  second moment
\bq\label{moment}
i \ne j \ \ \ ; \ \ \ \overline{M_{ij}M_{kn}} =
\fr{u^2}{(i-j)^2} (\delta_{jk} \delta_{in} + \delta_{jn}
\delta_{ik}) \ \ .
\ee
In general $u \sim 1/l_1$~. In the weak coupling limit ($U \ll t$) one
has $u \sim U l_1^{-1}$ (\ref{U2}) and in the strong coupling limit
($U \gg t$) one has $u \sim t^2 U^{-1} l_1^{-1}$ (\ref{U2str}). We
have omitted the slow dependence of $u$ on $i+j$
(\ref{U2},\ref{U2str}). This complication, if taken into account ,
should not change our main conclusions.  The diagonal elements of the
matrix $V$ are uniformly distributed within the range
\begin{eqnarray}\label{range}
 M_{nn}\equiv \varepsilon_n \ &,& \ 
 -\delta_E < \varepsilon_n < \delta_E \ \ , \\
\delta_E &=& Gu \ \ . \nonumber
\end{eqnarray}
Here we have subtracted from the total energy of each state the
trivial constant $E_0$ (\ref{delta}).  We have also introduced in
(\ref{range}) the dimensionless parameter $G=\delta_E /u$ which allows
to compare the strength of the diagonal and non-diagonal (\ref{moment})
elements of the matrix $M$. Analytical results in this section (as
well as the poor analytical estimates of the following one) will be
found only for $G\gg 1$. Moreover, just for $G\gg 1$ the
correspondence between $G$ and the parameters of the original
Hamiltonian (\ref{Ham}-\ref{Vint}) is most transparent. Due to the
duality between weak (\ref{me}) and strong (\ref{mestrong}) coupling
cases $G$ turns out to be the double-valued function of the strength
of interaction $U$
\begin{eqnarray}\label{GUt}
G=const\fr{t}{U}\ &,& \ 
{\rm \ for\ \ weak\ \ coupling\ } \ \ \ U\ll t
\ \ \ , \\
G=const\fr{U}{t}\ &,& \ 
{\rm for\ \ strong\ \ coupling} \ \ \ t\ll U
\ \ \ . \nonumber
\end{eqnarray}
The numerical constant in both cases is of the order of $1$. The
crossover from weak to strong coupling regime takes place at some
$G\sim 1$. Thus the matrix model (\ref{moment},\ref{range}) with very
small $G$ does not correspond to any physical limit of the Hamiltonian
(\ref{Ham}-\ref{Vint}). In the crossover region instead of the matrix
elements (\ref{me}) and (\ref{mestrong}) one should use some
renormalized new formula for the effective interaction like, e.g.,
(\ref{merenorm}). This effective interaction should account for the
strong ($\sim 100\%$ for $U\sim t$) renormalization of both
eqs.(\ref{me}) and (\ref{mestrong}) due to the regular mixing with
many two particle states which were not included into consideration by
the matrix model (\ref{moment},\ref{range}). Namely these are the
two-particle states with $ |\epsilon_{\scriptscriptstyle 1} +
\epsilon_{\scriptscriptstyle 2} - E_0| >\delta_E $ . However this 
renormalized interaction will naturally lead to the
same effective matrix model (\ref{moment},\ref{range}) with $G\sim 1$
(see also (\ref{merenorm}) and discussion around). Also $G$ is a
function of only $U$ and $t$ and at least for $w\ll U,t$ we do not
expect any considerable dependence of $G$ on the strength of disorder.

Matrices of the kind (\ref{moment}) are called Power-law Random Band
Matrices (PRBM).  Random Band Matrices have been the subject of
intensive investigation during the last years ( for review see
\cite{singapur}).  Power-law matrices with $M_{ij}\sim
1/|i-j|^\alpha$ also may be considered as a special type of random
band matrices. However our matrices (\ref{moment}) with $M_{ij}\sim
1/|i-j|$ play an outstanding role among other PRBM because just the
value $\alpha =1$ corresponds to a phase transition from localized to
delocalized eigenvectors \cite{mirlin}.

As we have tried to show in Section 2, due to the hierarchy of the
matrix elements (\ref{rowall}), the actual small parameter in our
problem, e.g., for the weak coupling case turns out to be not $U/t$
but $(U/t)\ln(l_1)$. The aim of this section will be to find
analytically $l_{ipr}$ (\ref{lipr}) for the case $(U/t) \ln(l_1) \gg 1
\gg U/t$ (or equivalently $\ln(l_1) \gg G \gg 1$). The method used for
this calculation follows the Renormalization Group approach of
ref. \cite{levitov}.

Thus we consider now the case $u\ll \delta_E \sim t/l_1$~. At a first
glance, in this case one has to treat the non-diagonal part of the
matrix $M$ (\ref{moment}) by perturbation theory. However the simple
perturbative estimate gives a very small correction $l_{ipr}-1 \sim
(u/\delta_E)^2$ to the inverse participation ratio. On the other hand,
no matter how weak is the interaction $u$ , with small $\sim
U/\delta_E$ probability the interval between the two diagonal elements
of $M$ may happen to be of the same order of magnitude as the
non-diagonal one: $\varepsilon_n -\varepsilon_k \sim M_{nk}$. These
rare events, which are not described by the usual perturbation theory,
turns out to be the most important for the chaotic mixing of original
non-perturbed eigen-vectors. Because the density of strongly
interacting pairs of levels is very low for $u\ll \delta_E$, one may
neglect in the zeroth approximation the effect of triple interaction,
as well as the more complicated events. Thus, it is enough to consider
the $2\times 2$ part $\tilde{M}$ of the full matrix $M$ with diagonal
matrix elements distributed uniformly between $-\delta_E$ and
$+\delta_E$, and the Gaussian distributed non-diagonal elements
$\tilde{M}_{1,2} = \tilde{M}_{2,1}=v$ with the mean squared variance
$\sigma^2 \ll \delta_E^2$ . The eigenvectors for this matrix are easy
to find
\begin{eqnarray}\label{meig}
&\tilde{M}& = \left( \begin{array}{c} \varepsilon_1 \ \  v \\
                                       v  \ \ \varepsilon_2
                                       \end{array} \right)
                                       \ \ ; \ \ 
\tilde{M}\left( \begin{array}{c} \alpha_1 \\
                                 \alpha_2
                                 \end{array} \right) =
\lambda \left( \begin{array}{c} \alpha_1 \\
                                \alpha_2
                                \end{array} \right)
                                \ \ ; \nonumber \\
&\alpha_{1,2}& = \sqrt{
\fr{2v^2}{\sqrt{(\varepsilon_1-\varepsilon_2)^2 + 4v^2}}  
\fr{1}{\sqrt{(\varepsilon_1-\varepsilon_2)^2 + 4v^2} \mp 
(\varepsilon_1
-\varepsilon_2)} } \ \ . 
\end{eqnarray}
The second solution is obtained by the permutation $\alpha_1
\rightarrow \alpha_2' \ , \ \alpha_2 \rightarrow - \alpha_1'$~.  The
participation ratio corresponding to the eigenstate (\ref{meig}) is 
\bq\label{pr}
P = l_{ipr}^{-1} = \alpha_1^4 +\alpha_2^4 = 1 -
\fr{2v^2}{(\varepsilon_1-\varepsilon_2)^2 + 4v^2} \ \ .
\ee
It is easy to find the averaged value: 
\bq\label{prav}
 \overline{P} = \int dv 
\fr{1}{\sqrt{2\pi}\sigma} \exp\left( - \fr{v^2}{2\sigma^2}
\right) \int_{-\delta_E}^{\delta_E} d\varepsilon_2
 \fr{r}{2\delta_E} =
1-\sqrt{\fr{\pi}{2}} \fr{\sigma}{\delta_E} \ \ .
\ee
This effect turns out to be much more important than the naive
perturbation theory estimate $1 - {P} \sim
(\sigma/\delta_E)^2$~.

Let us return to the $N\times N$ matrices (\ref{moment},\ref{range}).
In (\ref{prav}) we have found the averaged participation ratio for
each of the two eigenvectors of $\tilde{M}$ (\ref{meig}) caused by the
non-diagonal element $v$. Now each of the non-diagonal elements of the
full matrix $M_{mk}$ leads to the same correction to the participation
ratios of  two from the $N$ eigenvectors up to trivial replacement
$\sigma = u/|m-k|$. Therefore ($G=\delta_E/u$
(\ref{range}))
\bq\label{mpravall}
 \overline{P_2} =
1-\sqrt{\fr{\pi}{2}}\fr{2}{G}\sum_1^N \fr{1}{n} \approx 
1-\fr{\sqrt{2\pi}}{G} \ln(N) \ \ , \ \ 
\fr{1}{G} \ln(N) \ll 1 \ \ .
\ee
Because of $N\sim l_1$, we see again that our actual expansion
parameter is ${U}/{t} \ln(l_1)$ (weak coupling) in accordance with the
prediction of Section 2.

Physically the calculation of the participation ratio for our matrix
model resembles the calculation of the Partition function of the
interacting rare gas.  The subscript $2$ in $P_2$ in (\ref{mpravall})
means that we have taken into account only one double event
$\varepsilon_{n} -\varepsilon_{k} \sim M_{nk}$ and have completely
ignored all the triple collisions, four-particles collisions, and so
on. That is why the inequality describing the range of validity of
(\ref{mpravall}) is so restrictive.  However, let us consider now the
case when only $G^{-1}$ is small, but not $\ln(l_1)/G$. Consider the
triple event. Suppose that the three diagonal elements
$\varepsilon_i,\varepsilon_j,\varepsilon_k$ are close enough, so that
the basis vectors $e_i,e_j,e_k$ may be mixed strongly by the
interaction. Let us also have for definiteness
$|i-j|<|i-k|,|j-k|$. The crucial observation, which in fact allows to
find $l_{ipr}$ is that, due to $G\gg 1$, and the decrease of the
second moments (\ref{moment}) in the large $\ln(l_1)$ limit one may
strengthen the inequality:
\bq\label{ijk}
|i-j|<|i-k|,|j-k| \ \rightarrow \ |i-j|\ll |i-k|,|j-k| \ \ .
\ee
Now the calculation of the triple contribution to $P$ may be divided
into two stages. First, the vectors $e_i$ and $e_j$ are mixed by the
rather strong non-diagonal element $M_{ij}$ and form the two new basis
vectors $\tilde{e}_1$ and $\tilde{e}_2$ in accordance with
(\ref{meig})
\begin{eqnarray}\label{eeee}
\tilde{e}_1&=& \alpha_i e_i +\alpha_j e_j \ \ , \\
\tilde{e}_2&=& -\alpha_j e_i +\alpha_i e_j \ \ . \nonumber
\end{eqnarray}
After that, one of these states (say $\tilde{e}_1$) is mixed with $e_k$
by the much smaller elements $M_{ik}$ and $M_{jk}$. Due to
(\ref{moment}) and (\ref{ijk}), the variance of the effective matrix
element describing this mixing reads
\bq\label{perd}
\overline{|\langle \tilde{1} | M | k\rangle|^2} =
\overline{|\alpha_i M_{ik}+ \alpha_j M_{jk}|^2} =
\fr{u^2\alpha_i^2}{(i-k)^2} + \fr{u^2\alpha_j^2}{(j-k)^2}
\approx
\fr{u^2}{(i-k)^2} \approx \fr{u^2}{(j-k)^2} \ \ .
\end{equation}
Thus we see that the secondary interaction in the two stage triple
event (\ref{ijk}) effectively is not affected by the first
interaction.

This picture with the decoupling of the secondary interactions shows
the applicability of the Renormalization Group approach for the
calculation of $l_{ipr}$.  Our aim in fact is to diagonalize the
matrix $M$. Let us divide this procedure into $N$ steps. At the first
step let us diagonalize the three diagonal matrix
$\varepsilon_n,M_{n,n+1},M_{n,n-1}$. Because of $\varepsilon_n \gg
M_{n,n\pm 1}$ in general this procedure of diagonalization reduces
(approximately) to the simple nullification of the sub-diagonals
$M_{n,n\pm 1}$. Only for small part of states the diagonalization is
nontrivial. Suppose, that $\varepsilon_i-\varepsilon_{i+1} \sim
M_{i,i+1}$ for some $i$. Now for these two states $i$ and $i+1$ one
has to use the exact procedure (\ref{meig}). This diagonalization of
the $2\times 2$ sub-matrix should be followed by the orthogonal
transformation of the remaining part of the matrix $M$ : $M_{n,k>n+1},
M_{n,k<n-1}$. Namely one should perform the following transformation
of the two rows and two columns
\begin{eqnarray}\label{row}
\left( \begin{array}{ll}
M_{i,n}\\
M_{i+1,n}
       \end{array} \right)
&\rightarrow& 
\left( \begin{array}{ll}
\alpha_1 M_{i,n}+\alpha_2 M_{i+1,n}\\
-\alpha_2 M_{i,n}+\alpha_1 M_{i+1,n}
       \end{array} \right) \ \ \ , \\
\bigg( M_{n,i}\ ,\ M_{n,i+1}\bigg) 
&\rightarrow& 
\bigg( \alpha_1 M_{n,i}+\alpha_2 M_{n,i+1}\ , \
-\alpha_2 M_{n,i}+\alpha_1 M_{n,i+1}\bigg) \ \ , \nonumber
\end{eqnarray}      
where the index $n$ enumerates the elements of the $i$-th ($i+1$-th)
row or column.  $\alpha_1$ and $\alpha_2$ here are the same as in
(\ref{meig}) $|\alpha_1|^2+|\alpha_2 |^2 =1$. However, due to
(\ref{ijk}), the statistics of the matrix elements $M_{n,k>n+1},
M_{n,k<n-1}$ is not changed under the transformation (see
(\ref{perd})). In a similar way one may diagonalize out the third
diagonal $M_{n,n\pm 2}$, the fourth diagonal $M_{n,n\pm 3}$, and so
on.

Let us find, how the averaged participation ratio $\overline{P}$ flows
under such gradual diagonalization. It is convenient to write the
participation ratios of all $N$ states in one column $\{p\}$. Now the
gradual diagonalization of the matrix is accompanied by the gradual
change of $\{p\}$, which we illustrate by the formula:
\begin{eqnarray}\label{grp}
\left( \begin{array}{ll}
1\\
1\\
1\\
1\\
1\\
1\\
1\\
..
       \end{array} \right)
\rightarrow  
\left( \begin{array}{ll}
1\\
p_{\scriptscriptstyle 1}\\
p_{\scriptscriptstyle 1}\\
1\\
1\\
1\\
1\\
..
       \end{array} \right)
\rightarrow 
\left( \begin{array}{ll}
1\\
p_{\scriptscriptstyle 1}\\
p_{\scriptscriptstyle 1}\\
1\\
p_{\scriptscriptstyle 2}\\
1\\
p_{\scriptscriptstyle 2}\\
..
       \end{array} \right)
\rightarrow 
\left( \begin{array}{ll}
1\\
p_{\scriptscriptstyle 1}p_{\scriptscriptstyle 3}\\
p_{\scriptscriptstyle 1}\\
1\\
p_{\scriptscriptstyle 2}p_{\scriptscriptstyle 3}\\
1\\
p_{\scriptscriptstyle 2}\\
..
       \end{array} \right)
\rightarrow 
\left( \begin{array}{ll}
p_{\scriptscriptstyle 4}\\
p_{\scriptscriptstyle 1}p_{\scriptscriptstyle 3}\\
p_{\scriptscriptstyle 1}\\
1\\
p_{\scriptscriptstyle 2}p_{\scriptscriptstyle 3}
p_{\scriptscriptstyle 4}\\
1\\
p_{\scriptscriptstyle 2}\\
..
       \end{array} \right)
\rightarrow 
\ ... \ \ , 
\end{eqnarray}
where the $p_{\scriptscriptstyle k}$'s are the participation ratios
for the sufficiently rare ``important'' events happening at the $k$-th
step\cite{rigorous}. The probability to have the
``important'' event of the amplitude $p$ is described by
(\ref{pr},\ref{prav}) and the $p_{\scriptscriptstyle k}$ at each step
are uniformly distributed over the column $\{p\}$. Combining together
(\ref{mpravall}) and (\ref{grp}), one finds:
\begin{eqnarray}\label{otvet}
 \overline{P} = {\fr{1}{l_{ipr}}} &=& \prod_{n=1}^N 
 \left( 1-\fr{\sqrt{2\pi}}{nG} \right)
 \sim N^{\displaystyle -\gamma_{ipr}}
 \sim l_1^{\displaystyle -\gamma_{ipr}} 
 \ \ , \\
\gamma_{ipr} &=& \fr{\sqrt{2\pi}}{G} \ \ . \nonumber
\end{eqnarray}
Of course this result was found for $G\gg 1$. Nevertheless the number
of chaotically mixed states due to (\ref{otvet}) may be very large
$l_{ipr}\gg 1$. Most probably for $G$ not large the formula
$l_{ipr}\sim l_1^{\displaystyle \gamma_{ipr}}$ would be still valid,
although $\gamma_{ipr}$ in this case will depend on $G$ in a more
complicated way than in (\ref{otvet}). At least our numerical results
which will be considered in the following section support this
conjecture.  It is also possible to extend the renormalization group
approach of \cite{levitov} used in this section in order to look for
$\gamma_{ipr}$ (\ref{otvet}) as a series in $1/G$.

\section{Effective matrix model. Large sample}\label{sec:6}

Now at last we are going to consider our main problem: the coherent
propagation of two interacting particles in a very large disordered
sample. In order to get the adequate matrix model we have to combine
the block picture considered at the end of Section 2 with our
understanding of the hierarchy of the matrix elements
(\ref{es2}),(\ref{U2}),(\ref{U2str}). Any pair of overlapping
single-particle wave-functions may be associated with one block from
the row of blocks described in (\ref{block}). As in the case of the short
sample, considered in the previous section, it seems to be enough to
take into account only the chaotic mixing of the states from the
narrow zone (\ref{delta}). Thus again there are effectively $\sim l_1$
states in each block. It is natural to take into account only the
interaction of the pairs belonging to the neighboring
blocks. Therefore the exact Hamiltonian (\ref{Ham}) may be replaced by
the block-wise random matrix
\begin{eqnarray}\label{mwise}
M=
\left( \begin{array}{ll}
M_1^b, \tilde{M_1^b},\ \: 0\ ,\ \: 0\ ,..\\
\tilde{M_1^b},M_2^b, \tilde{M_2^b},\ \: 0\ ,..\\
\ \: 0\ ,\tilde{M_2^b},M_3^b, \tilde{M_3^b},..\\
\ \: 0\ ,\ \: 0\ ,\tilde{M_3^b},M_4^b,..\\
\ .\ .\ .\ .\ .\ .\ .\ .\ .\ .\ .\ .\
       \end{array} \right) \ \ .
\end{eqnarray}
Here $M_i^b$ and $\tilde{M_i^b}$ are the random $l_b\times l_b$ (note
that $l_b\sim l_1$) matrices describing the transitions within $i$-th
and between $i$-th and $i+1$-th block respectively.  In general the
total size of the matrix ($\sim L$) is supposed to be much larger than
all the other lengths which may appear in our problem.

In order to exploit the hierarchy of the matrix elements of the
interaction one has to enumerate properly the simple states within one
block as it was done for the short sample in the previous section:
$n(\epsilon_1) > n'(\epsilon_1')$ if $\epsilon_1 > \epsilon_1'$. After
that statistics of the diagonal block matrices $M_i^b$ will be
again described by the formulas (\ref{moment}),(\ref{range}). Moreover,
the non-diagonal square blocks $\tilde{M_i^b}$ from (\ref{mwise}) will
be also described by (\ref{moment}) (of course the non-diagonal blocks
$\tilde{M}$ have no strong diagonal (\ref{range})).

It is convenient to ``smooth out'' the block-wise matrix (\ref{mwise})
so that
\begin{eqnarray}\label{smoothm}
 M_{nn}= \varepsilon_n \ \ &,& \ \ 
 -G < \varepsilon_n < G \ \ , \\
0<|i-j|<2l_b  \ \ &,&  \ \ \overline{M_{ij}M_{kn}} = 
 (\delta_{jk} \delta_{in} + \delta_{jn}
\delta_{ik}) F(i-j) \ \ ,\nonumber\\
{\rm where}  &\,&  F(i-j)= 
\left(\fr{a}{(i-j)^2}+\fr{1}{(i-j-l_b)^2}+
\fr{1}{(i-j+l_b)^2}\right) \ \ ,\nonumber\\
2l_b \le |i-j| \ \ &,&  \ \ M_{ij}=0 \ \ . \nonumber
\end{eqnarray}
Here we use the dimensionless parameter $G$ introduced in
(\ref{range}). The connection between $G$ and the parameters of the
original Hamiltonian (\ref{Ham},\ref{Vint}) is described by
(\ref{GUt}).  In order to go from (\ref{mwise}) to (\ref{smoothm}) we
have changed only a small fraction of the matrix elements which in the
large logarithmic case $\ln(l_1)\gg 1$ are not important for the
coherent propagation length. The amplitude $a$ of the central peak of
$F$ in (\ref{smoothm}) is of the order of one: $a\sim 1$ ($a>1$).

The elements of the matrix (\ref{smoothm}) vanish, if $|i-j| > const
\times l_1$. Therefore, as for usual band matrices its eigen-vectors
should have some finite localization length. Moreover, because of the
size of the block in (\ref{mwise}) is $l_b\sim l_1$ this localization
length measured in vector indices should be proportional to the
coherent propagation length $l_2$ measured in the sites of original
lattice (\ref{H1}). Thus the quasi-$1d$ matrix Hamiltonian
(\ref{smoothm}) accounts for both the dimension of our sample and the
complicated structure of the matrix elements of interaction
(\ref{es2}),(\ref{U2}),(\ref{U2str}). The central peak of the function
$F(i-j)$ (\ref{smoothm}) has the same form as the second moment
(\ref{moment}) for the matrix model considered before for the
single-block sample. However, as we will see below just the satellite
peaks $\sim (i-j\pm l_b)^{-2}$ in (\ref{smoothm}) are mainly
responsible for the large localization length. Therefore, for future
discussion and for numerical simulations we have further simplified
the model (\ref{smoothm}), so that
\begin{eqnarray}\label{smoothn}
 M_{nn}= \varepsilon_n \ \ &,& \ \ 
 -G < \varepsilon_n < G \ \ , \\
i<j\le i+l_1  \ \ &,&  \ \ \overline{M_{ij}M_{kn}} = 
 \fr{\delta_{jk} \delta_{in} + \delta_{jn}
\delta_{ik}}{(j-i-l_1+1)^2} \ \ ,\nonumber\\
i+l_1 < j \ \ &,&  \ \ M_{ij}=0  \ \ ,\nonumber\\
&\,&  M_{ji}=M_{ij}\ \ .\nonumber
\end{eqnarray}
This matrix model looks quite similar with that considered in the
previous section (\ref{moment}),(\ref{range}). The main difference is
that in (\ref{moment}) the elements of the matrix decrease in the
power law fashion as one goes farther from the diagonal, while in
(\ref{smoothn}) the matrix elements reaches the maximal amplitude at
the two sub-diagonals $j=i\pm l_1$ and decrease like the power of
$|i-j\pm l_1|$ towards the main diagonal. The model (\ref{smoothn})
should reproduce all the important features of the (\ref{smoothm}) and
thus will lead to qualitatively correct description of coherent
propagation.

The band matrices with Cauchy distribution of the matrix elements were
also considered in \cite{MMM}. This model due to the power law tail of 
the probability to find the large matrix element looks much closer to 
our (\ref{smoothn}) than the usually considered SBRM matrix models
\cite{Dima,Fyodorov,Frahm,Pichard3}. However, as is confirmed also by 
the author of \cite{MMM}, for a rigorous application to the TIP 
problem more rigorous investigations are required.

For $G\gg 1$ one may try to use the approach of Section 5 in order to
find at least the inverse participation ratio $l_{ipr}$ for the model
(\ref{smoothn}). Moreover, if $\ln(l_1)\ll G$ only the rare double
interactions of the energy levels should be taken into account and the
models (\ref{moment}) and (\ref{smoothn}) became equivalent. In
particular the result (\ref{mpravall}) still holds for the matrices
(\ref{smoothn}) up to trivial replacement $N\rightarrow l_1$. However,
taking into account of even the triple events for the model
(\ref{smoothn}), turns out to be much more complicated problem than
for the model (\ref{moment}).

First of all, the Renormalization Group so successful for the PRBM of
the previous section does not work for the model (\ref{smoothn}). One
may try again to perform the gradual diagonalization
(\ref{row}),(\ref{grp}) for the matrices (\ref{smoothn}). However, now
this procedure should start from the most important sub-diagonal
$M_{i,i\pm (l_1-1)}$. As a result, the orthogonal transformation
(\ref{row}) in this case will change significantly  statistics of
the remaining part of the matrix. For example, if $\varepsilon_i
-\varepsilon_{i+l} \sim M_{i,i+l}$, one has for two elements from the
``remaining part''
\begin{eqnarray}\label{rowl}
\left( \begin{array}{ll}
M_{i+1,i+l}\\
M_{i+1,i}
       \end{array} \right)
\rightarrow
\left( \begin{array}{ll}
\alpha_1 M_{i+1,i+l}\\
-\alpha_2 M_{i+1,i+l}
       \end{array} \right) \ \ \ , 
\end{eqnarray}      
because of $M_{i+1,i} \ll M_{i+1,i+l}$ due to the (\ref{smoothn}).
Suppose for example that $\alpha_1 = \alpha_2=1/\sqrt{2}$. This means
that due to the equation (\ref{rowl}) instead of one large matrix
element $M_{i+1,i+l}$ we have obtained two of them. Each of these two
matrix elements is $\sqrt{2}$ times smaller than the original one, but
their sum which in fact is important (see
(\ref{prav}),(\ref{mpravall})) is $\sqrt{2}$ larger. Thus effectively
the secondary interactions in (\ref{smoothn}) are enhanced compared to
those for the PRBM.

Moreover, the accurate calculation of the triple contribution to the
participation ratio for the model (\ref{smoothn}) gives
\bq\label{pirper}
 \overline{P} =
1-\sqrt{2\pi}\fr{1}{G_{eff}} - 
const \left(\fr{1}{G_{eff}}\right)^2 
\ln G_{eff} - ...  \ \ , \ \ 
G _{eff}= \fr{G}{\ln(l_1)} \ \ .
\ee
For short we skip the calculation of the $const \sim 1$. One sees, that
the triple collision contribution compared to the PRBM result
(\ref{otvet}) is enhanced by the additional factor $\ln(G_{eff})$. In
order to find this logarithm one has to consider the effective matrix
element of the two stage interaction
\bq\label{chanels}
\langle i | M_{eff} | j \rangle = \sum_k 
\fr{M_{ik}M_{kj}}{\varepsilon_j-\varepsilon_k} \ \ .
\ee
The contribution of such matrix element to the participation ratio
(see (\ref{pr},\ref{prav})) is proportional to $\delta P\sim -|\langle
i | M_{eff} | j \rangle|$ and the averaging over intermediate energy
$\varepsilon_k$ naturally gives the logarithm. As one may show for
usual PRBM this logarithm does not appear due to the competition
between the effective second order matrix element (\ref{chanels}) and
the direct one $M_{ij}$.

The higher order corrections to (\ref{pirper}) may be also shown to
have the form $const\times G_{eff}^{-n-1} (\ln
G_{eff})^n$. Unfortunately, the explicit summation of these logarithmic
corrections, even if possible, will hardly teach us anything. On the
one hand, if $G_{eff}^{-1} \ln(G_{eff})\ll 1$ the $l_{ipr}$ is also
small. On the other hand, the interesting case $G_{eff}^{-1}
\ln(G_{eff})\sim 1$ corresponds to $\ln(G_{eff})\sim 1$ also and the
large logarithm approximation (\ref{pirper}) does not work in this
case. Thus, the only conclusion we may draw about the inverse
participation ratio and moreover the localization length for the model
(\ref{smoothn}) is that they are some unknown functions of
$G_{eff}^{-1}=\ln(l_1)/G$ (\ref{pirper}).

Finally, we have the convincing explanation why the chaotic mixing of
many states for the model (\ref{smoothn}) (as well as
(\ref{mwise}),(\ref{smoothm})) should be enhanced compared to that for
the PRBM model (\ref{moment}),(\ref{range}). However, we have no idea
how even to estimate analytically the magnitude of this
mixing. Therefore, the last possibility we have in hands is to perform
the numerical simulations.

The main limitation for our numerical procedure is the
impossibility to diagonalize the very large random matrices. Namely,
all our data below have been found for matrices of the size $L=100 -
1000$ or sometimes $L<2500$ (we use the notation $L$ for the total
size of the matrix in order to stress that it is directly associated
with the length of our sample). In order not to take into account the
boundary effects (for $i\sim l_1$ and $L-i\sim l_1$, where $i$ is the
vector index) we have considered the coherent propagation in the
periodic $1d$-sample. The periodic generalization of the model
(\ref{smoothn}) reads
\begin{eqnarray}\label{mper} 
 -G <  &M_{nn}&  < G \ \ , \\
i\ne j   \ &,&  \ \overline{M_{ij}M_{kn}} = 
 (\delta_{jk} \delta_{in} + \delta_{jn}
\delta_{ik}) \bigg[ F(i-j) +F(i-j-L) +F(i-j+L)\bigg] \ ,\nonumber\\
&\,& F(x)= \left\{ \begin{array}{ll}
(l_1-|x|)^{-2} & {\rm if} \ \ 0<|x|<l_1 \\
     0         & {\rm otherwise}
     \end{array} \right.
 \ \ .\nonumber
\end{eqnarray}
For comparison of the results for inverse participation ratio
$l_{ipr}$ with those for more simple model of the previous section we
have also performed some calculations for the same periodic matrices
(\ref{mper}) but with
\bq\label{mperepem}
F(x)=x^{-2} \ \ , \ \ 0<|x|<l_1 \ \ .
\ee

\begin{figure}
\epsfxsize=11 truecm
\centerline{\epsffile{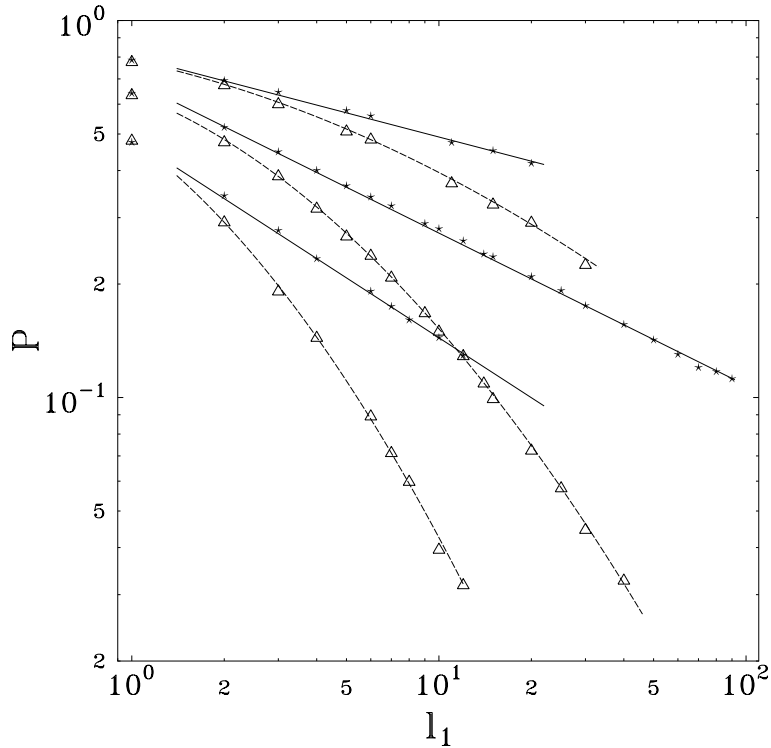}}
\caption{ Participation ratio $\overline P $ for the model (77)
- triangles and the simplified power law model (78) -
stars. The upper, middle and bottom curves corresponds to different
 strengths of the diagonal $G=10.,5.,2.5$ . }
\end{figure}

The participation ratio $\overline{P}$ averaged over different
eigenvectors as well as over many ($10-50$) realizations of the random
matrix (\ref{mper},\ref{mperepem}) is shown in the fig. 1 as a
function of $l_1$. It is natural to use the logarithmic scale for both
$\overline{P}$ and $l_1$. We have performed the calculations for three
values of the strength of the diagonal $G=2.5,5.,10.$. In order to get
rid of the edge effects close to the border of the energy zone we have
taken into account only the vectors with eigenvalue $|\varepsilon_i|<
0.8 G$ for $G=5.,10.$ and $|\varepsilon_i|< 0.7 G$ for $G=2.5$. The
first point $l_1 = 1$ coincides for both models as may be seen from
(\ref{mper},\ref{mperepem}). However we are interested in the
case $l_1 \gg 1$.

The few comments on the figure 1 are of order.
\begin{itemize}
\item The log-log curves for the model (\ref{mperepem}) (solid lines)
shows no deviation from the straight line in agreement with
(\ref{otvet}).
\item The slope of both curves for the realistic model (\ref{mper})
and for the toy one (\ref{mperepem}) grows with $G$. This is again in
agreement with (\ref{otvet}) and with what one may expect from, e.g.,
(\ref{pirper}), but disagrees with what one should expect for the
model of \cite{Dima}.
\item The participation ratio for realistic model (\ref{mper}) decays
for large $l_1$ much faster than those for (\ref{mperepem}) in
accordance with our wave-handing discussion
(\ref{pirper},\ref{chanels}) of the role of satellite peaks in
(\ref{smoothm}).
\item The dependence of $\ln(\overline{P})$ on $\ln(l_1)$ for the
model (\ref{mper}) is not linear at least for $l_1<25$ as may be seen
from the figure. We have fitted these data by a parabola (dashed
lines). However, our computations are not enough to say with sure
either say $\overline{P}\sim \exp(-const\times \ln(l_1)^2)$ for large
$l_1$ or asymptotically $\overline{P}\sim l_1^{ -const}$ and the
finite curvature of the dashed lines in the fig. 1 accounts only for
the small $l_1$ effects?
\end{itemize}

The calculation of $\overline{P}$ was carried out straightforwardly in
accordance with the definition (\ref{lipr},\ref{pr}). The calculation
of the localization length for circular matrices due to the
identification of indices $i+L=i$ requires a little more care. Let us
consider our circular vector index $i$ literally and define the radius
of the ``center of mass'' for the eigenvector
\begin{eqnarray}\label{cmass}
 R \! &=& \! \sqrt{\overline{x}^2 + \overline{y}^2} 
\ \ , \\
&\,& \overline{x}= \sum_{i=1}^{L} |\alpha_i|^2
\cos\left(\fr{2\pi i}{L}\right) \ \ , \ \ 
\overline{y}= \sum_{i=1}^{L} |\alpha_i|^2
\sin\left(\fr{2\pi i}{L}\right) \ \ , \nonumber
\end{eqnarray}
where $\alpha_i$ are the normalized eigenvector components. This $ R $
naturally reduces to $ R =1$ for the very localized (pointlike) states
and to $\langle R \rangle=0$ for the delocalized ones. One may easily
see that for very large matrices the deviation of $\langle R \rangle$
from unit is proportional to the mean squared size of the wave packet
$l_2^2=\langle (i-i_0)^2 \rangle$. Therefore for explicit definition
of $l_2$ we have used the formula
\bq\label{blb}
l_2 = \sqrt{\langle (i-\langle i \rangle)^2 \rangle}
=\fr{L}{\pi}\sqrt{\fr{1-\langle R \rangle}{2}} 
\left[ 1+ O\left(\fr{l_2^{\,2}}{L^2}\right)\right] \ \ .
\ee
Here the brackets $\langle ... \rangle$ stands for averaging over the
many eigenvectors as well as many realizations of the random matrix
(\ref{mper}). In practice we have done a few runs of computations for
each $l_1$ with different values of the overall size of the matrices
$L$ in order to fit out the $\sim (l_2/L)^2$ finite size
corrections. The finite size effects for $l_{ipr}$ (fig. 1) are
naturally much weaker $\sim \exp(-const \ L/l_2)$.
 
The enhancement factors for the coherent propagation $l_2/l_1$ found
for three different values of the strength of the diagonal
$G=2.5,5.,10.$ are shown in the figure 2. It is to be noted that $G$
measures the relative strength of the interaction
(\ref{Ham})-(\ref{Vint}): $G\sim U/t$ for weak coupling and $G\sim
t/U$ for strong coupling.

\begin{figure}[t]
\epsfxsize=11 truecm
\centerline{\epsffile{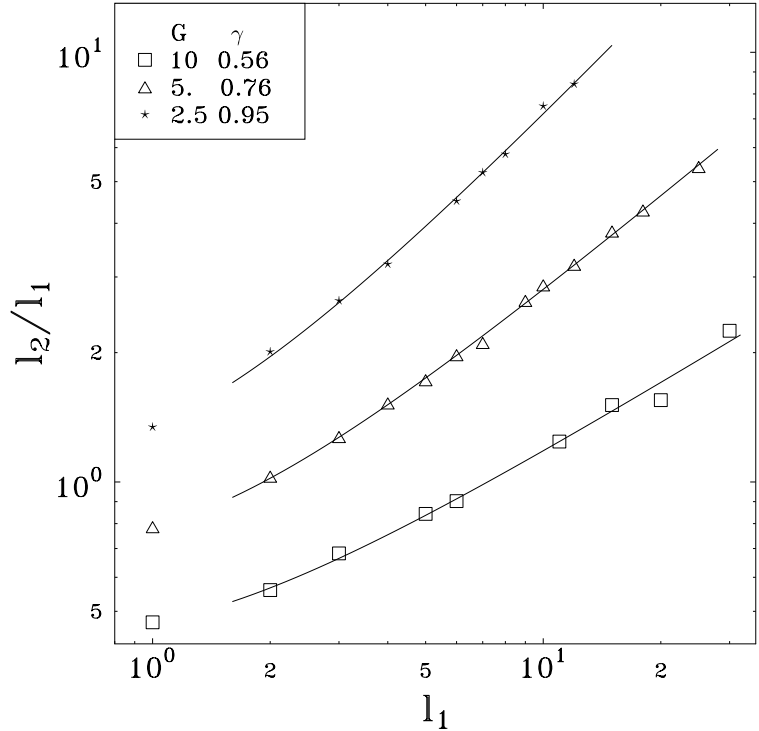}}
\caption{The factor of enhancement for the coherent propagation
$l_2/l_1$ as a function of $l_1$ for few values of $G$. Both axes
given in logarithmic scale. The $\gamma$ is the slope of the curve at
the asymptotics $l_1\rightarrow \infty$ (81).}
\end{figure}

As one can see from the fig. 2 the deviations from linear behavior in
the logarithm--logarithm plot for the $l_2/l_1$ turn out to be 
sufficiently weaker
than for the participation ratio $P$. The solid curves on the figure
corresponds to the least squares fitting with the formula
\begin{eqnarray}\label{fit}
&\,&
y=a+\gamma x + c \exp(-x) \ \ \ , \\
&\,&
y=\ln(l_2/l_1) \ \ ,  \ \  x=\ln(l_1) \ \ . \nonumber
\end{eqnarray}
This approximation corresponds to
\bq\label{lll}
l_2/l_1 \sim l_1^{\displaystyle
\gamma}\left( 1+ \fr{c}{l_1}\right) \ \ .
\ee
Thus, we have conjectured that the preassymptotic corrections decrease
like $l_1^{-1}$. The values of the exponent $\gamma$ for different
magnitude of the diagonal $G$ are
\begin{eqnarray}\label{GGG}
&\,&
G=10. \ \ \ \ \gamma=0.56\, \pm 0.04  \ \ ,  \nonumber \\
&\,&
G=5. \, \ \ \ \ \ \gamma=0.76\, \pm 0.04\ \ ,   \\
&\,&
G=2.5  \ \ \ \ \gamma=0.95\, \pm 0.04 \ \ ,  \nonumber 
\end{eqnarray}
In fact these three $\gamma$ may be considered as a main result of our
paper.  The pure statistical errors for this $\gamma$-s are of the
order of a few percent and may be further improved easily. The main
physical problem however is the proper choice of the fitting function
(\ref{fit}). For example the data presented in the fig. 2 still allows
one to use $y=a+\gamma x+cx^2$ instead of (\ref{fit}).

First of all, from the result (\ref{GGG}) one may conclude with sure
that the exponent $\gamma$ depends on the strength of the diagonal
$G$. However, with only three points in hand we were afraid to fit the
dependence $\gamma (G)$ by some smooth function. The physical
conditions, evidently consistent with the data (\ref{GGG}), for the
function $\gamma (G)$ are $\gamma (G=\infty)=0$ and $\gamma
(G=0)=const \sim 1$. Although the case $G\ll 1$ does not correspond to
any mapping of the original TIP problem as we have said after the
eq. (\ref{GUt}). Also the connection between $G$, and $U$ and $t$ is
described in eq. (\ref{GUt}) and below.

The only rigorous way to confirm numerically the equation (\ref{lll})
and to exclude the more exotic dependence of $l_2$ on $l_1$ is to
consider larger $l_1$. On the other hand, in using of the formula
(\ref{blb}) we assume that the total size of the sample is large
compared to the coherent propagation length $L\gg l_2$. Therefore
below we would like to develop the method which in principle allows
one to consider the coherent propagation with only the small matrices
$L\sim l_2$ or even $L\ll l_2$ in hand.

Let us define the new function 
\bq\label{cret}
F(L,l_1)\equiv \langle \sqrt{1-R(L,l_1)} \ \rangle
 = F(L/l_2) \ \ \ .
\ee
Here the last equality will be our scaling hypothesis. Due to
(\ref{cmass}),(\ref{blb}) $F=1$ for the very small sample $L\ll l_2$
and $F \sim l_2/L $ for $L\gg l_2$. For $L\sim l_2$ the deviation of
$F$ from trivial $F=1$ will measure the typical warp of the wave
function on the circle. One should expect that this typical
inhomogeneity of the wave function as well as the averaged shift of
center of mass $\langle R(L,l_1) \rangle$ will depend simply on the
ratio of the size of the sample $L$ and the two-particle localization
length $l_2$. Thus, it is natural to suppose that for $L\gg l_1$ the
function $F(x)$ is universal in the sense that all dependence on $l_1$
in the r.h.s. of (\ref{cret}) is hidden in $l_2(l_1)$. One still may
have some doubts in this quite natural conjecture due to, e.g., the
multifractal nature of the wave function. However, we may easily
control its validity within our numerical calculations.

One may fix the value of $F$ and solve numerically the
equation for $L$
\begin{eqnarray}\label{creti}
&\,&
F(L_c,l_1)= F(x_c)=F_c \ \ \ , \\
&\,&
L_c=x_c(F_c)\times l_2(l_1) \ \ \ . \nonumber
\end{eqnarray}
Here only the overall normalization factor $x_c$ depends on the value
of $F_c$. Of course, the scaling behavior (\ref{cret}),(\ref{creti})
should be violated at $L\sim l_1$. However, because  we have seen from 
fig. 2 that for large single particle localization length $l_2 \gg l_1
$ one may hope that (\ref{creti}) will be still valid even for $l_1\ll
L \ll l_2$. Physically, this means that we are trying to find the
manifestations of the finite coherent propagation length $l_2$ in the
sample much smaller than this length.

\begin{figure}[t]
\epsfxsize=12 truecm
\centerline{\epsffile{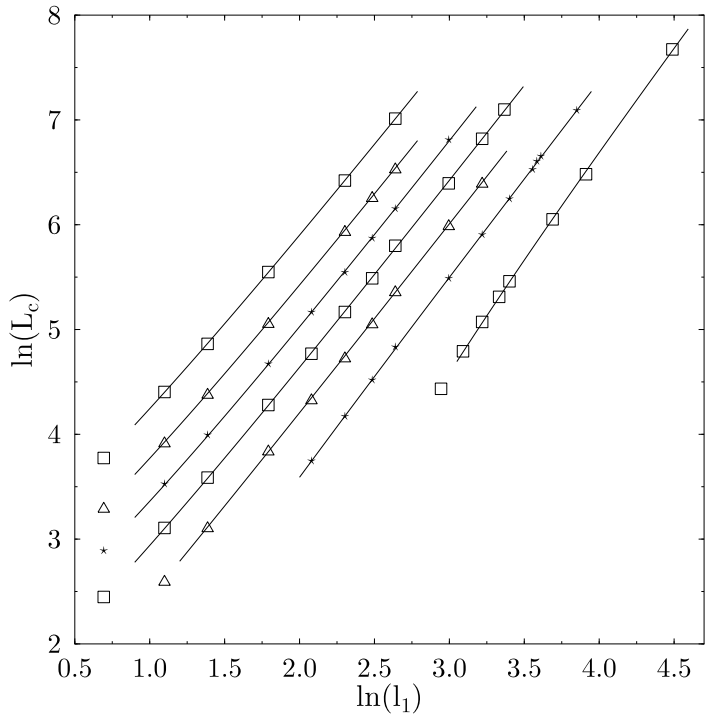}}
\caption{The $\ln(L_c)$ as a function of $\ln(l_1)$
(84),(85) for $G=5$ and various values of $F_c= 0.2,
0.3, 0.4, 0.5, 0.6, 0.7, 0.8$ (from the left to the right).}
\end{figure}

The dependence of the $\ln(L_c)$ on the $\ln(l_1)$ for $G=5.$ and for
different values $F_c= 0.2, 0.3, 0.4, 0.5, 0.6, 0.7, 0.8$ is presented
in the fig. 3. At least six of this seven curves looks quite parallel
which may be considered as the confirmation of the scaling hypothesis
(\ref{cret}). Also it is natural that the corrections to scaling as it
may be seen from the figure are stronger for larger values of
$F_c$. In order to take into account at least the corrections to
scaling proportional to $\sim 1/l_1$ we like in (\ref{fit}) have
fitted our results by
\begin{eqnarray}\label{cretin}
&\,&
y_i=a_i+(1+\gamma) x + c_i \exp(-x) \ \ \ , \\
&\,&
y_i=\ln(L_c(F_c)) \ \ ,  \ \  x=\ln(l_1) \ \ , \nonumber
\end{eqnarray}
where $a_i$ and $c_i$ differs for different $F_c$ but $\gamma$ is the
same for all seven curves of fig. 3. Finally the joint fitting
(\ref{cretin}) gives
\bq\label{cretino}
\gamma =0.83 \ \pm 0.04\ \ \ {\rm for} \ \ \ \ G=5. \ \ ,
\ee
in agreement with (\ref{GGG}). The main advantage of this
(\ref{cret})-(\ref{cretin}) indirect method of determination of
$\gamma$ is that we have reached the value $l_1=90$ ($l_2\approx
1300$) by considering the $L\times L$ matrices with $L<1500$. For the
direct method (\ref{blb})-(\ref{lll}) for $G=5.$ in order to reach
even $l_1=25$ we have to invoke the much larger matrices with $L\sim
2500$.

Of course, due to the logarithmic scale the progress made by going
from fig. 2 to fig. 3 may look not so impressive and the result
(\ref{cretino}) due to the scaling hypothesis (\ref{cret}) may be
considered as model dependent. Nevertheless we conclude from
(\ref{cretino}) that we see no evidence for the violation of the
simple power law behavior (\ref{lll}) for $l_2$.

\section{Conclusions}\label{sec:7}

In the present paper we have tried to revise the issue of the
enhancement of coherent propagation of two interacting particles in
random potential predicted by D.L.~Shepelyansky \cite{Dima}. 

First of all, we definitely see the enhancement but it is of the
different nature and has the different functional dependence from
those usually considered. The existence of the enhancement itself is
proved by considering the analytical estimate of the matrix element of
interaction potential
(\ref{es2}),(\ref{U2}),(\ref{U2str}). Effectively the
particle-particle interaction turns out to be enhanced in logarithm of
the Anderson localization length times $\ln(l_1)$. Moreover, after
summation of high order contributions in this effective interaction
this logarithm is most likely exponentiated so that $l_2/l_1 \sim
\exp\{ \gamma \ln(l_1)\}$~, where $\gamma =\gamma(U/t)$~.

Unfortunately the functional dependence of the two particle
localization length (\ref{lll}) was found only numerically. Therefore
we still can not completely exclude some other, rather exotic,
dependence of $l_2$ on $l_1$. The equation (\ref{lll}) finds also some
support in the very similar behavior of the inverse participation
ratio $l_{ipr}$ (\ref{otvet}) which was found analytically for the
simplified model with the short sample.

The results (\ref{lll}),(\ref{otvet}) were obtained by mapping of the
original two-particle problem onto some random matrix model
(\ref{moment}),(\ref{range}),(\ref{lll}),(\ref{smoothn}). This means
that we are able to explain only qualitatively the behavior of
$l_2$. Therefore it is quite desirable if somebody will found the same
behavior of the coherent propagation length as in our equation
(\ref{lll}) in the direct calculation with the Hamiltonian (\ref{Ham})
(some numerical results supporting (\ref{lll}) may be found in
\cite{Pichard}). To this end the numerical method of ref. \cite{Oppen}
seems to be very useful. The method of investigation of the effect of
localization in the samples not large compared to the localization
length described at the end of Section 6 being combined with the
numerical method of \cite{Oppen} may also allow one to consider the
larger range of variation of the coherent propagation length.

In general, may be the most interesting result of our paper is that we
have found the nontrivial structure (or hierarchy of the elements) of
the matrix of interaction (\ref{es2}),(\ref{U2}),(\ref{U2str}) in the
basis of noninteracting states
(\ref{pbasis}),(\ref{symm}),(\ref{spbasis}). Just due to this
hierarchy the matrix models we have to consider differ so drastically
from those investigated before.

In this paper we consider only the interacting particles moving in
exactly $1d$-random potential. It will be very interesting to
generalize our approach for interacting particles in weak two and
three dimensional random potential, which are currently the subject of
intensive investigation \cite{AGKLB}.

{\bf Acknowledgments.}  Authors are thankful to V.~F.~Dmitriev,
Y.~Imry, I.~V.~Kolokolov, A.~D.~Mirlin, D.~V.~Savin, V.~V.~Sokolov,
V.~B.~Telitsin, V.~G.~Zelevinsky and especially to O.~P.~Sushkov and
D.~L.~Shepelyansky for numerous helpful discussions.


\begin{thebibliography}{99}

\bibitem{Dima} D.L.~Shepelyansky, Phys. Rev. Lett. {\bf 73}, 2607
(1994).


\bibitem{Imry} 
 Y.~Imry, Europhys.~Lett~{\bf 30}, 405 (1995).

\bibitem{Pichard} 
 K.~Frahm, A.~M\"uller-Groeling, J.-L.~Pichard and 
D.~Weinmann, Europhys.~Lett. ~{\bf 31}, 405 (1995).

\bibitem{Pichard2} 
D.~Weinmann, A.~M\"uller-Groeling, J.-L.~Pichard and K.~Frahm, 
Phys. Rev. Lett. ~{\bf 75}, 1598 (1995).

\bibitem{Oppen} F.~von~Oppen, T.~Wettig and J.~M\"uller,
Phys.~Rev.~Lett.  ~{\bf 76}, 491 (1996)  \\ The actual enhancement
parameter observed in this paper is rather weak $l_2/l_1\approx
3$. However, within the definition of $l_2$ used by authors the limit
of switched of interaction corresponds to $l_2/l_1=1/2$. Therefore
effectively the observed enhancement is twice larger.


\bibitem{Dima3} 
 P.~Jacquod and D. L.~Shepelyansky, Phys.~Rev.~Lett. ~{\bf 75}, 
3501 (1995).

\bibitem{Fyodorov} Y.~V.~Fyodorov and A.~D.~Mirlin, Phys.~Rev. B ~{\bf
 52}, R11580 (1995).

\bibitem{Frahm} K.~Frahm and A.~M\"uller-Groeling,
 Europhys.~Lett. ~{\bf 32}, 385 (1995).

\bibitem{Pichard3} K.~Frahm, A.~M\"uller-Groeling, and J.-L.~Pichard,
Phys.~Rev.~Lett.  ~{\bf 76}, 1509 (1996).

\bibitem{Pichard4} D.~Weinmann and J.-L.~Pichard, Phys.~Rev.~Lett.
~{\bf 77}, 1556 (1996).

\bibitem{Dima4} F.~Borgonovi and D. L.~Shepelyansky, Journal de
 Physique I.  {\bf 6}(2) (1996) 287.

\bibitem{Sushkov} P.~Jacquod, D.~L.~Shepelyansky and O.~P.~Sushkov,
cond-mat/9605141. See however, the discussion at the end of our
Section 2.

\bibitem{Dorokhov}
{The propagation of interacting
particles in the random potential was considered also in: \\
O.~N.~Dorokhov, Sov. Phys.-JETP {\bf 71}(2) (1990) 360.\\
However, the confining particle--particle interaction
considered by Dorokhov seems to be very different from the short
range interaction of ref. \cite{Dima}.}

\bibitem{molecules} In the presence of, e.g., attractive interparticle
interaction there evidently appear a few trivial molecular bound
states. The size of this molecules is $x_1-x_2 \sim \lambda$ and the
``molecular'' localization length $\Delta (x_1+x_2) \sim l_1$.  More
concretely, the size of the molecule for weak interaction (if the size
of the molecule is larger than the lattice spacing) is $x_1-x_2 \sim
{t/U}$ (\ref{H1},\ref{Vint}). However, for very small $U$ one should
also take into account the decay of the molecules due to interaction
with the random potential.  Anyway the number of molecular states is
$\Delta N_m\sim L/\lambda$ -- much less than the number of coherent
states.

\bibitem{Pichthanks} One of the natural ways to investigate the effect
of interaction is to consider the rigidity of spectrum for TIP as it
was done in \cite{Pichard4}. In accordance with our consideration of
the large $U$ limit the authors of ref. \cite{Pichard4} have observed
the decrease of level repulsion for the strong interaction case $U>t$
as well as for the weak interaction one $U<t$. We are thankful to
J.-L.~Pichard for discussion of this result.

\bibitem{Pastur} I.~M.~Lifshits, S.~A.~Gredeskul, L.~A.~Pastur,
Introduction to the theory of disordered systems. John Wiley and Sons,
Inc., 1988.

\bibitem{levitov} L.~Levitov, Europhys. Lett. {\bf 9}, 83 (1989);
Phys. Rev. Lett. {\bf 64}, 547 (1990).

\bibitem{mirlin} A.~D.~Mirlin, Y.~V.~Fyodorov, F.~Dittes, J.~Quezada,
and T.~H.~Seligman, Phys.~Rev.~E, {\bf 54}, 3221 (1996)

\bibitem{discussion} The authors
of ref. \cite{Oppen} have fitted their results for any strength of 
the interaction by (\ref{l2new}) with only one fixed value 
$\gamma \equiv 1$. This fixed value of $\gamma$ is the most important
difference of the result of \cite{Oppen} from our prediction. 
Because the large $U$ were not
considered in \cite{Oppen} we would expect the most clear deviation
from $\gamma = const$ in their data for $U\ll t$. In the notations of
\cite{Oppen} the small $U$ are hidden in the small argument $u\xi_1/t$
part of their fig. 1. However, just the small argument part of this
figure demonstrates the largest deviation from the straight line
expected by the authors. On the other hand, the center of the zone
used for numerical investigation in \cite{Oppen} turns out to be just
the most difficult case for the theoretical analysis. For example, the
weak coupling limit does not work at all here because in the center of
the zone energy conservation automatically leads to 
the momentum
conservation ($v_1=v_2$ in (\ref{U2}) see also discussion after the
eq. (\ref{U2str})).

\bibitem{Thouless} D.~C.~Thouless,  Phys.~Rev.~Lett.
~{\bf 39}, 1167 (1977).

\bibitem{MacKinnon} See e.g. B.~Kramer and A.~MacKinnon,
Rep. Prog. Phys. {\bf 56}, 1469 (1993).

\bibitem{singapur} Y.~V.~Fyodorov, A.~D.~Mirlin,
Int. J. Mod. Phys. {\bf B8}, 3795 (1994).

\bibitem{rigorous}{To be more rigorous at this point one should introduce
some $\varepsilon \ll 1$. If the relative change of the participation
ratio for the $i$-th level at the $k$-th step is less than
$\varepsilon$, this correction is ignored. The ``important'' events
are those with $p_k<1-\varepsilon$.}

\bibitem{MMM} D.~L.~Shepelyansky, cond-mat/9603086.

\bibitem{AGKLB} Ya.~M.~Blanter, cond-mat/9604101;\\
B.~L.~Altshuler, Y.~Gefen, A.~Kamenev, and
L.~S.~Levitov, cond-mat/9609132.

\end{thebibliography}
\end{document}